\documentclass[useAMS,usenatbib,usegraphicx]{mn2e}
\usepackage{times}
\usepackage{epsfig}
\title[Reddening, Colour and Metallicity of M31 GCs]
{Reddening, Colour and Metallicity of the M31 Globular Cluster
System}

\author[Z. Fan et al.]
{Z. Fan$^{1,2}$, J. Ma$^1$,\thanks{E-mail: majun@bac.pku.edu.cn}
R. de Grijs$^{3,1}$ and X. Zhou$^1$\\
$^1$National Astronomical Observatories, Chinese Academy of
Sciences, 20A Datun Road, Chaoyang District, Beijing 100012,
China\\
$^2$Graduate University of Chinese Academy of Sciences,
19A Yuquan Road, Shijingshan District, Beijing 100049, China\\
$^3$Department of Physics \& Astronomy, The University of
Sheffield, Hicks Building, Hounsfield Road, Sheffield S3 7RH}

\date{Received; Accepted}

\pubyear{2007}

\begin{document}

\label{firstpage}

\maketitle

\begin{abstract}
Using metallicities from the literature, combined with the Revised
Bologna Catalogue of photometric data for M31 clusters and cluster
candidates (the latter of which is the most comprehensive catalogue
of M31 clusters currently available, including 337 confirmed
globular clusters -- GCs -- and 688 GC candidates), we determine 443
reddening values and intrinsic colours, and 209 metallicities for
individual clusters without spectroscopic observations. This, the
largest sample of M31 GCs presently available, is then used to
analyse the metallicity distribution of M31 GCs, which is bimodal
with peaks at $\rm {[Fe/H]}\approx -1.7$ and $-0.7$ dex. An
exploration of metallicities as a function of radius from the M31
centre shows a metallicity gradient for the metal-poor GCs, but no
such gradient for the metal-rich GCs. Our results show that the
metal-rich clusters appear as a centrally concentrated spatial
distribution; however, the metal-poor clusters tend to be less
spatially concentrated. There is no correlation between luminosity
and metallicity among the M31 sample clusters, which indicates that
self-enrichment is indeed unimportant for cluster formation in
M31.\\ The reddening distribution shows that slightly more than half
of the GCs are affected by a reddening of $E(B-V) \la 0.2$ mag; the
mean reddening value is $E(B-V) = 0.28_{-0.14}^{+0.23}$ mag. The
spatial distribution of the reddening values indicates that the
reddening on the northwestern side of the M31 disc is more
significant than that on the southeastern side, which is consistent
with the conclusion that the northwestern side in nearer to us.

\end{abstract}


\section{Introduction}
\label{Introduction.sec}

The formation and evolution scenarios of the Milky Way Galaxy still
remain among the most important unsolved problems in contemporary
astrophysics \citep{per02}. One promising way for us to better
understand, and to possibly make progress towards addressing these
issues is by studying globular clusters (GCs). GCs are generally
considered the fossils of the galactic formation and evolution
processes, since they form during the very early stages of their
host galaxy's evolution \citep{bh00}. GCs are very dense,
gravitationally bound spherical systems of several thousands to more
than a million stars. They can be observed out to much greater
distances than individual stars, so that they can be used, and are
in fact the ideal tracers, to study the properties of extragalactic
systems. The most distant GC systems that have been studied to date
are those in the Coma cluster; for instance, \citet{baum95}
presented GC counts in the bright elliptical galaxy NGC 4881
\citep[at a distance of $\simeq 108$ Mpc;][]{baum95} based on {\sl
Hubble Space Telescope (HST)}/WFPC2 observations. Using data of
similar quality, \citet{Kavelaars00}, \citet{harris00} and
\citet{wh00} published a series of papers on the GCs in NGC~4874 and
IC 4051, the central cD galaxy and a giant elliptical galaxy in the
Coma cluster, respectively.

Located at a distance of about 780 kpc \citep{sg98, mac01}, M31 is
the nearest and largest spiral galaxy in the Local Group of
galaxies. It contains 337 confirmed GCs and 688 GC candidates
\citep{gall04}, thus allowing us access to a much larger number of
GCs than in our own Galaxy. However, despite the difference in GC
numbers, from the observational evidence collected so far
\citep[see, e.g.,] []{rich05}, the M31 GCs and their Galactic
counterparts reveal some striking similarities
\citep{ffp94,dj97,bhh02}. More recently, \citet{kim07} embarked on a
new systematic wide-field CCD survey of M31 GCs, and found 1164 GCs
and GC candidates -- of which 559 are previously known GCs and 605
newly-found GC candidates; based on survey data from the Canada
France Hawaii telescope and Wide Field Camera on the Isaac Newton
telescope, \citet{huxor07} combined his detailed analysis of the M31
GC system with recent results based on the galaxy's stellar halo,
and concluded that M31 and the Milky Way are rather more similar
than previously thought. Therefore, studying the properties of the
GCs in M31 may not only improve our understanding of the formation
and structure of our nearest large neighbour galaxy, but also that
of our own Galaxy. However, there are also significant
differences\footnote{Huxor (2007) suggests that the primary
difference between the Galaxy and M31, and between their GC systems
in particular, is likely due to the more vigorous recent merger
history of M31.} between the GC systems of the Milky Way and M31: in
particular, M31 has a much larger population of GCs than the Milky
Way \citep[see] [and references therein]{kim07}; there are
populations of ``faint fuzzies'' and extended GCs in the outer halo
of M31 that are not seen in the Milky Way \citep[]
[]{huxor05,mackey06,mackey07}, and there is a population of young to
intermediate-age GCs in M31 that are (again) not seen in the Milky
Way again \citep[see, eg.,] []{beasley04,puzia05}.

A reliable estimate of the reddening, caused by dust contamination, is
important for the study of the stellar population of a given GC, in
order to obtain its intrinsic spectral energy distribution. In
general, on galaxy-wide scales dust tends to be distributed close the
the galactic plane in galactic discs. Therefore, the disc clusters are
most affected by extinction due to dust in the galactic disc (in
addition to the effects due to Galactic foreground extinction). A
reliable estimate of the extinction caused by Galactic material along
a given line of sight can be obtained from the reddening maps of
\citet{bh82} and \citet{Schlegel98}. However, a reliable estimate of
the internal extinction in a given galaxy is not easy to obtain. More
specifically, although there were some ways in which to deal with the
problem of determining the reddening toward the GCs in M31 prior to
the publication of \citet{bh00}, some may not have been fully
satisfactory. For example, \citet{van69} assumed a uniform reddening
for the clusters when he pioneered integrated-light spectroscopy of
M31 GCs, whereas \citet{ir85} assumed GCs to have a single intrinsic
colour when using them as reddening probes; some authors \citep[see
e.g.][]{km60,ve62,har74,van75,bg77} assumed that the halo GCs in M31
were only affected by Galactic foreground extinction, based on which
they then averaged the intrinsic colours of the halo GCs and
subtracted this value from the observed colours to obtain the
reddening values for all the GCs \citep[see details in][]{bh00}.
There are also two other ways to determine the reddening of M31
clusters, which seem reasonable but they have only been applied to a
handful of clusters: \citet{Frogel80} estimated the individual
reddening for 35 clusters using the reddening-free parameter $Q_K$,
based on unpublished spectroscopic data by L. Searle;
\citet{Crampton85} used the intrinsic colours of the 35 clusters
obtained by \citet{Frogel80} to calibrate $(B-V)_{\rm 0}$ as a
function of spectroscopic slope parameter $S$ of the continuum between
$\sim$ 4000 and 5000 \AA, and then determined the intrinsic colours
for about 40 GCs and GC candidates.

\citet{bh00} presented a new catalogue of photometric and
spectroscopic data of M31 GCs, and determined the reddening for each
individual cluster using correlations between optical and infrared
colours and metallicity, and by defining various ``reddening-free''
parameters based on this catalogue. \citet{bh00} found that the M31
and Galactic GC extinction laws (see their table 6), and the M31 and
Galactic GC colour-metallicity relations are similar to each
other. They then estimated the reddening to M31 objects with
spectroscopic data using the relation between intrinsic optical
colours and metallicity as determined for Galactic clusters. For
objects without spectroscopic data, they used the relationships
between the reddening-free parameters and certain intrinsic colours,
based on the Galactic GC data. \citet{bh00} compared their results
with those in the literature and confirmed that their estimated
reddening values are reasonable, and quantitatively consistent with
previous determinations for GCs across the entire M31 disc. In
particular, \citet{bh00} showed that the distribution of reddening
values as a function of position appears reasonable in that the
objects with the smallest reddening are spread across the disc and
halo, while the objects with the largest reddening are concentrated in
the galactic disc. The reddening values for M31 clusters obtained by
\citet{bh00} are widely used \citep[see, e.g.,][]{jiang03,Ma06b,
fan06,rey06}

To study the metal abundance properties of GCs can help us understand
the formation and enrichment processes of their host galaxy. For
example, if galaxies form as a consequence of a monolithic,
dissipative and rapid collapse of a single massive, nearly-spherical
spinning gas cloud in which the enrichment time-scale is shorter than
the collapse time, the halo stars and GCs should show large-scale
metallicity gradients \citep{eggen62,bh00}. On the other hand,
\citet{sz78} presented a chaotic scheme for the early evolution of a
galaxy, in which loosely bound pre-enriched fragments merge with the
main body of the proto-galaxy over a significant period, in which case
there should be a more homogeneous metallicity distribution. At
present, most galaxies are thought to have formed through a
combination of both of these scenarios (see also Section 4.5)

{\sl HST} provides a unique tool for studying GCs in external
galaxies. For example, based on data from the {\sl HST} archive,
\citet{Gebhardt99}, \citet{Larsen01} and \citet{kw01} showed that many
large galaxies possess two or more subpopulations of GCs that have
quite different chemical compositions \citep[see also][]{west04}.
Recently, \citet{peng06} presented the colour distributions of GC
systems for 100 early-type galaxies from the ACS Virgo Cluster Survey,
and found that, on average, galaxies at all luminosities appear to
have bimodal or asymmetric GC colour/metallicity distributions. The
presence of colour bimodality among these old GCs indicates that there
have been at least two major star-forming mechanisms in the (early)
histories of massive galaxies \citep{west04,peng06,strader06}.

Based on the newest photometric and spectral data, in this paper we
determine reliable reddening values for 443 GCs and GC candidates (the
largest GC sample in M31 used to date), and we also determine the
metallicities for 209 GCs and GC candidates without spectroscopic
observations. We then perform a statistical analysis using this GC
sample. We describe the photometric and spectroscopic data for the M31
GCs in Section \ref{data.sec}. In Section \ref{red.sec}, we determine
and analyse the reddening of our M31 GC sample. Section
\ref{metal.sec} is devoted to our statistical analysis. Finally, the
discussion and conclusions of this paper are presented in Section
\ref{Conclu.sec}.

\section{Database}
\label{data.sec}

\subsection{The sample}

\citet{gall04} present the final Revised Bologna Catalogue of M31 GCs
including 337 confirmed GCs and 688 GC candidates which compose our
primary sample. From a comparison with \citet{bh00} and \citet{per02},
89 candidates turn out not to be GCs, and these were thus removed from
the sample. More recently, \citet{huxor07} provided a further revision
of the Bologna Catalogue, including a number of additional clusters in
the halo of M31. Because he only provides photometry for these new
clusters in two filters, we will not include these in our final
sample, however (see below regarding the photometric requirements of
the method we adopt in this paper).

\subsection{The optical and near-infrared photometric data}

The source of the photometric data utilised in this paper is from the
catalogue of \citet{gall04}, i.e. the updated Bologna Catalogue with
homogenised optical ($UBVRI$) photometry collected from the most
recent photometric references available in the literature. In this
catalogue, \citet{gall04} used the $UBVRI$ photometry from
\citet{bh00} as a reference to obtain the master catalogue of
photometric measurements as homogeneously as possible. In addition,
\citet{gall04} identified 693 known and candidate GCs in M31 using the
2MASS database, and included their 2MASS $JHK_{\rm s}$ magnitudes,
transformed to the CIT photometric system \citep{Elias82,Elias83}.
\citet{gall04} compile a final table including the $UBVRIJHK_{\rm s}$
photometric data for the 337 confirmed and 688 candidate GCs in M31
(their table 2), which is the photometric database we use in this
paper.

\subsection{Spectroscopic metallicities}
\label{Metal.sec}

\citet{hbk91} obtained spectroscopy of 150 M31 GCs with the Multiple
Mirror Telescope (MMT). The system they used has a resolution of 8--9
{\AA} and enhanced blue sensitivity. Their observations extend to the
atmospheric cut-off at 3200 {\AA}, since many of the strongest and
most metallicity-sensitive spectral features of interest are in the UV
\citep[see details in] []{bh90}. \citet{hbk91} determined the
metallicities for these 150 GCs using six absorption-line indices from
integrated cluster spectra, following \citet{bh90}.

\citet{bh00} determined the metallicities of 61 M31 GCs and GC
candidates using the Keck Low Resolution Imaging Spectrometer (LRIS)
and the MMT Blue Channel spectrograph. They computed the
absorption-line indices using the method of \citet{bh90}. \citet{bh00}
combined the measured indices based on the metallicity calibration
from \citet{bh90}, in order to determine metallicities. Their
metallicities were found to be consistent with those of
\citet{hbk91}. Finally, \citet{bh00} compiled a catalogue of
spectroscopic metallicities for 188 M31 GCs
\footnote{http://cfa-www.harvard.edu/$\sim$huchra/m31globulars/m31glob.dat}.

\citet{per02} determined the metallicities of 201 GCs in M31 using the
Wide-Field Fibre Optic Spectrograph (WYFFOS) at the 4.2m William
Herschel Telescope. They calculated 12 absorption-line indices
following the method of \citet{bh90}. By comparison of the line
indices with the previously published M31 GC [Fe/H] values of
\citet{bonoli87}, \citet{bh90}, and \citet{bh00}, \citet{per02} found
that the line indices of the CH (G band), Mg~$b$ and Fe53 lines best
represent the metal abundances of their observed targets. They
determined the final metallicities of these GCs based on an unweighted
mean of the [Fe/H] values calculated from the G band, Mg~$b$, and Fe53
line strengths.

There are some GCs in our sample for which the metallicities were
determined in two or three of the studies mentioned above. To use all
the metallicities as coherently as possible, we use the metallicities
of \citet{per02} whenever available, since \citet{per02} present the
largest (homogeneous) sample of M31 GC metallicities. For the
metallicities determined by \citet{hbk91} and \citet{bh00}, there is
only one GC in common, object B328.  \citet{hbk91} and \citet{bh00}
determined its metallicity at {$\rm [Fe/H]=-1.22\pm0.80$} and {$\rm
[Fe/H]=-1.51\pm0.28$}, respectively.  We adopt the metallicity from
\citet{bh00}, given its smaller associated uncertainty.

Overall, we obtained metallicities for 295 M31 GCs, which we list in
Table \ref{t1.tab}. We will refer to these data as our spectroscopic
metallicity catalogue (hereafter SMCat). We will use the SMCat to
perform our statistical analysis.

\citet{bh00} and \citet{per02} determined the GC metallicities using
the metallicity calibration defined in \citet{bh90}. Therefore, all
three metallicity determinations are on the same [Fe/H] system.
\citet[][their fig. 7]{per02} show convincingly that there are no
systematic offsets among these three sets of metallicity
determinations.

\begin{table*}
\caption{Spectroscopic metallicities of the M31 GCs collected in this
paper.} \label{t1.tab}
\begin{center}
\begin{tabular}{crccrccrccrc}
\hline
\hline
Name & \multicolumn{1}{c}{$\rm [Fe/H]$} &   source & Name &  \multicolumn{1}{c}{$\rm [Fe/H]$} &   source & Name & \multicolumn{1}{c}{$\rm [Fe/H]$} &  source & Name &  \multicolumn{1}{c}{$\rm [Fe/H]$} &  source \\
\hline

G055    & $ -1.07\pm 0.55$ &  3&B011    & $ -1.54\pm 0.34$ &  3&G001    & $ -1.08\pm 0.09$ &  3&G002    & $ -1.70\pm 0.36$ &  3\\
B009    & $ -1.57\pm 0.26$ &  3&B020    & $ -1.07\pm 0.10$ &  3&B023    & $ -0.92\pm 0.10$ &  3&B024    & $ -0.48\pm 0.30$ &  3\\
B027    & $ -1.64\pm 0.32$ &  3&B044    & $ -1.14\pm 0.37$ &  3&B046    & $ -1.84\pm 0.61$ &  3&B058    & $ -1.45\pm 0.24$ &  3\\
B063    & $ -0.87\pm 0.33$ &  3&B064    & $ -1.55\pm 0.30$ &  3&B068    & $ -0.29\pm 0.59$ &  3&B073    & $ -0.64\pm 0.46$ &  3\\
B085    & $ -1.83\pm 0.40$ &  3&B086    & $ -1.74\pm 0.17$ &  3&B092    & $ -1.65\pm 0.49$ &  3&B095    & $ -1.57\pm 0.41$ &  3\\
B096    & $ -0.26\pm 0.43$ &  3&B098    & $ -0.67\pm 0.58$ &  3&B103    & $ -0.56\pm 0.62$ &  3&B106    & $ -0.86\pm 0.68$ &  3\\
B107    & $ -1.18\pm 0.30$ &  3&B112    & $  0.29\pm 0.49$ &  3&B115    & $ -0.15\pm 0.38$ &  3&B131    & $ -0.81\pm 0.28$ &  3\\
B143    & $  0.09\pm 0.42$ &  3&B146    & $ -0.43\pm 0.81$ &  3&B151    & $ -0.75\pm 0.18$ &  3&B152    & $ -0.87\pm 0.49$ &  3\\
B153    & $ -0.08\pm 0.33$ &  3&B154    & $ -0.45\pm 0.63$ &  3&B163    & $ -0.36\pm 0.27$ &  3&B165    & $ -1.80\pm 0.32$ &  3\\
B174    & $ -1.67\pm 0.27$ &  3&B178    & $ -1.51\pm 0.12$ &  3&B183    & $ -0.19\pm 0.31$ &  3&B201    & $ -1.06\pm 0.21$ &  3\\
B205    & $ -1.34\pm 0.13$ &  3&B206    & $ -1.45\pm 0.10$ &  3&B211    & $ -1.67\pm 0.52$ &  3&B212    & $ -1.75\pm 0.13$ &  3\\
B228    & $ -0.65\pm 0.66$ &  3&B229    & $ -1.81\pm 0.74$ &  3&B233    & $ -1.59\pm 0.32$ &  3&B239    & $ -1.18\pm 0.61$ &  3\\
B240    & $ -1.76\pm 0.18$ &  3&B317    & $ -2.12\pm 0.36$ &  3&B318    & $ -2.10\pm 0.50$ &  3&B343    & $ -1.49\pm 0.17$ &  3\\
B344    & $ -1.13\pm 0.21$ &  3&B352    & $ -1.88\pm 0.83$ &  3&B357    & $ -0.80\pm 0.42$ &  3&B358    & $ -1.83\pm 0.22$ &  3\\
B373    & $ -0.50\pm 0.22$ &  3&B375    & $ -1.23\pm 0.22$ &  3&B376    & $ -2.18\pm 0.99$ &  3&B377    & $ -2.19\pm 0.65$ &  3\\
B379    & $ -0.70\pm 0.35$ &  3&B381    & $ -1.22\pm 0.43$ &  3&B384    & $ -0.66\pm 0.22$ &  3&B387    & $ -1.96\pm 0.29$ &  3\\
B397    & $ -1.05\pm 0.53$ &  3&B403    & $ -0.45\pm 0.78$ &  3&B405    & $ -1.80\pm 0.31$ &  3&B407    & $ -0.85\pm 0.33$ &  3\\
B430    & $ -1.80\pm 0.65$ &  3&B431    & $ -1.23\pm 0.57$ &  3&B486    & $ -2.28\pm 0.98$ &  3&NB16    & $ -1.36\pm 0.12$ &  3\\
NB61    & $  0.26\pm 0.57$ &  3&NB65    & $ -0.78\pm 0.52$ &  3&B036    & $ -0.99\pm 0.25$ &  2&B126    & $ -1.20\pm 0.47$ &  2\\
B292    & $ -1.42\pm 0.16$ &  2&B302    & $ -1.50\pm 0.12$ &  2&B304    & $ -1.32\pm 0.22$ &  2&B310    & $ -1.43\pm 0.28$ &  2\\
B328    & $ -1.51\pm 0.28$ &  2&B337    & $ -1.09\pm 0.32$ &  2&B350    & $ -1.47\pm 0.17$ &  2&B354    & $ -1.46\pm 0.38$ &  2\\
B383    & $ -0.48\pm 0.20$ &  2&B401    & $ -1.75\pm 0.29$ &  2&NB67    & $ -1.43\pm 0.13$ &  2&NB68    & $ -0.76\pm 0.33$ &  2\\
NB74    & $ -0.02\pm 0.43$ &  2&NB81    & $ -0.75\pm 0.33$ &  2&NB83    & $ -1.26\pm 0.16$ &  2&NB87    & $  0.26\pm 0.41$ &  2\\
NB89    & $ -0.53\pm 0.57$ &  2&NB91    & $ -0.71\pm 0.33$ &  2&B295    & $ -1.71\pm 0.15$ &  1&B298    & $ -2.07\pm 0.11$ &  1\\
B301    & $ -0.76\pm 0.25$ &  1&B303    & $ -2.09\pm 0.41$ &  1&DAO023  & $ -0.43\pm 0.13$ &  1&B305    & $ -0.90\pm 0.61$ &  1\\
B306    & $ -0.85\pm 0.71$ &  1&DAO025  & $ -1.96\pm 0.97$ &  1&B307    & $ -0.41\pm 0.36$ &  1&B311    & $ -1.96\pm 0.07$ &  1\\
B312    & $ -1.41\pm 0.08$ &  1&B314    & $ -1.61\pm 0.32$ &  1&B313    & $ -1.09\pm 0.10$ &  1&B315    & $ -2.35\pm 0.54$ &  1\\
B001    & $ -0.58\pm 0.18$ &  1&DAO030  & $ -0.65\pm 0.34$ &  1&B316    & $ -1.47\pm 0.23$ &  1&B319    & $ -2.27\pm 0.47$ &  1\\
B321    & $ -2.39\pm 0.41$ &  1&G047    & $ -1.19\pm 0.29$ &  1&B004    & $ -0.31\pm 0.74$ &  1&B005    & $ -1.18\pm 0.17$ &  1\\
B443    & $ -2.37\pm 0.46$ &  1&B327    & $ -2.33\pm 0.49$ &  1&B006    & $ -0.58\pm 0.10$ &  1&B195D   & $ -1.64\pm 0.19$ &  1\\
B008    & $ -0.41\pm 0.38$ &  1&B010    & $ -1.77\pm 0.14$ &  1&B012    & $ -1.65\pm 0.19$ &  1&B448    & $ -2.16\pm 0.19$ &  1\\
DAO036  & $ -2.16\pm 0.32$ &  1&B013    & $ -1.01\pm 0.49$ &  1&B335    & $ -1.05\pm 0.26$ &  1&B015    & $ -0.35\pm 0.96$ &  1\\
B016    & $ -0.78\pm 0.19$ &  1&B451    & $ -2.13\pm 0.43$ &  1&B017    & $ -0.42\pm 0.45$ &  1&B018    & $ -1.63\pm 0.77$ &  1\\
DAO039  & $ -1.22\pm 0.41$ &  1&B019    & $ -1.09\pm 0.02$ &  1&B021    & $ -0.90\pm 0.06$ &  1&B338    & $ -1.46\pm 0.12$ &  1\\
DAO041  & $ -1.14\pm 0.30$ &  1&B453    & $ -2.09\pm 0.53$ &  1&B341    & $ -1.17\pm 0.05$ &  1&V031    & $ -1.59\pm 0.06$ &  1\\
B025    & $ -1.46\pm 0.13$ &  1&B026    & $  0.01\pm 0.38$ &  1&B028    & $ -1.87\pm 0.29$ &  1&B029    & $ -0.32\pm 0.14$ &  1\\
B030    & $ -0.39\pm 0.36$ &  1&B031    & $ -1.22\pm 0.40$ &  1&B342    & $ -1.62\pm 0.02$ &  1&B033    & $ -1.33\pm 0.24$ &  1\\
B034    & $ -1.01\pm 0.22$ &  1&DAO047  & $ -1.13\pm 0.57$ &  1&B035    & $ -0.20\pm 0.54$ &  1&V216    & $ -1.15\pm 0.26$ &  1\\
B037    & $ -1.07\pm 0.20$ &  1&B038    & $ -1.66\pm 0.44$ &  1&B039    & $ -0.70\pm 0.32$ &  1&B040    & $ -0.98\pm 0.48$ &  1\\
DAO048  & $ -2.01\pm 0.99$ &  1&B041    & $ -1.22\pm 0.23$ &  1&B042    & $ -0.78\pm 0.31$ &  1&B043    & $ -2.42\pm 0.51$ &  1\\
B045    & $ -1.05\pm 0.25$ &  1&B458    & $ -1.18\pm 0.67$ &  1&B047    & $ -1.62\pm 0.41$ &  1&B048    & $ -0.40\pm 0.37$ &  1\\
B049    & $ -2.14\pm 0.55$ &  1&B050    & $ -1.42\pm 0.37$ &  1&B051    & $ -1.00\pm 0.13$ &  1&B053    & $ -0.33\pm 0.26$ &  1\\
B054    & $ -0.45\pm 0.17$ &  1&B055    & $ -0.23\pm 0.07$ &  1&B056    & $ -0.06\pm 0.10$ &  1&B057    & $ -2.12\pm 0.32$ &  1\\
B059    & $ -1.36\pm 0.52$ &  1&B061    & $ -0.73\pm 0.28$ &  1&B065    & $ -1.56\pm 0.03$ &  1&B066    & $ -2.10\pm 0.35$ &  1\\
B069    & $ -1.35\pm 0.43$ &  1&V246    & $ -1.35\pm 0.29$ &  1&B072    & $ -0.38\pm 0.25$ &  1&B074    & $ -1.88\pm 0.06$ &  1\\
B075    & $ -1.03\pm 0.33$ &  1&B076    & $ -0.72\pm 0.06$ &  1&B081    & $ -1.74\pm 0.40$ &  1&B082    & $ -0.80\pm 0.18$ &  1\\
B083    & $ -1.18\pm 0.44$ &  1&B088    & $ -1.81\pm 0.06$ &  1&B090    & $ -1.39\pm 0.80$ &  1&B091    & $ -1.80\pm 0.61$ &  1\\
B093    & $ -1.03\pm 0.12$ &  1&NB20    & $ -0.80\pm 0.23$ &  1&B094    & $ -0.17\pm 0.45$ &  1&NB33    & $  0.04\pm 0.38$ &  1\\
B097    & $ -1.21\pm 0.13$ &  1&B102    & $ -1.57\pm 0.10$ &  1&B105    & $ -1.13\pm 0.32$ &  1&B109    & $ -0.13\pm 0.41$ &  1\\
B110    & $ -1.06\pm 0.12$ &  1&B116    & $ -0.88\pm 0.12$ &  1&B117    & $ -1.33\pm 0.45$ &  1&B119    & $ -0.49\pm 0.18$ &  1\\
B122    & $ -1.69\pm 0.34$ &  1&B125    & $ -1.52\pm 0.08$ &  1&B127    & $ -0.80\pm 0.14$ &  1&B129    & $ -1.21\pm 0.32$ &  1\\
B130    & $ -1.28\pm 0.19$ &  1&B134    & $ -0.64\pm 0.08$ &  1&B135    & $ -1.62\pm 0.04$ &  1&B355    & $ -1.62\pm 0.43$ &  1\\
B137    & $ -1.21\pm 0.29$ &  1&B140    & $ -0.88\pm 0.77$ &  1&B141    & $ -1.59\pm 0.21$ &  1&B144    & $ -0.64\pm 0.21$ &  1\\
DAO058  & $ -0.87\pm 0.07$ &  1&B148    & $ -1.15\pm 0.34$ &  1&B149    & $ -1.35\pm 0.25$ &  1&B467    & $ -2.49\pm 0.47$ &  1\\
B356    & $ -1.46\pm 0.28$ &  1&B156    & $ -1.51\pm 0.38$ &  1&B158    & $ -1.02\pm 0.02$ &  1&B159    & $ -1.58\pm 0.41$ &  1\\
B160    & $ -1.17\pm 1.25$ &  1&B161    & $ -1.60\pm 0.48$ &  1&B164    & $ -0.09\pm 0.40$ &  1&B166    & $ -1.33\pm 0.37$ &  1\\
B167    & $ -0.42\pm 0.23$ &  1&B170    & $ -0.54\pm 0.24$ &  1&B272    & $ -1.25\pm 0.16$ &  1&B171    & $ -0.41\pm 0.04$ &  1\\
B176    & $ -1.60\pm 0.10$ &  1&B179    & $ -1.10\pm 0.02$ &  1&B180    & $ -1.19\pm 0.07$ &  1&B182    & $ -1.24\pm 0.12$ &  1\\
B185    & $ -0.76\pm 0.08$ &  1&B184    & $ -0.37\pm 0.40$ &  1&B188    & $ -1.51\pm 0.17$ &  1&B190    & $ -1.03\pm 0.09$ &  1\\
B193    & $ -0.44\pm 0.17$ &  1&G245    & $ -0.31\pm 0.16$ &  1&B472    & $ -1.45\pm 0.02$ &  1&B197    & $ -0.43\pm 0.36$ &  1\\
B199    & $ -1.59\pm 0.11$ &  1&B198    & $ -1.13\pm 0.30$ &  1&B200    & $ -0.91\pm 0.61$ &  1&B203    & $ -0.90\pm 0.32$ &  1\\
\hline
\end{tabular}
\end{center}
\end{table*}
\addtocounter{table}{-1}

\begin{table*}
\caption{Continued.}
\label{t2.tab}
\begin{center}
\begin{tabular}{crccrccrccrc}
\hline
\hline Name & \multicolumn{1}{c}{$\rm [Fe/H]$} &   source & Name &  \multicolumn{1}{c}{$\rm [Fe/H]$} &   source & Name & \multicolumn{1}{c}{$\rm [Fe/H]$} &  source & Name &  \multicolumn{1}{c}{$\rm [Fe/H]$} &  source \\
\hline
B204    & $ -0.80\pm 0.17$ &  1&B207    & $ -0.81\pm 0.59$ &  1&B208    & $ -0.84\pm 0.04$ &  1&B209    & $ -1.37\pm 0.13$ &  1\\
B210    & $ -1.90\pm 0.32$ &  1&B213    & $ -1.02\pm 0.11$ &  1&B214    & $ -1.00\pm 0.61$ &  1&DAO065  & $ -1.80\pm 0.36$ &  1\\
DAO066  & $ -1.82\pm 0.26$ &  1&B216    & $ -1.87\pm 0.39$ &  1&B217    & $ -0.93\pm 0.14$ &  1&B218    & $ -1.19\pm 0.07$ &  1\\
B219    & $ -0.01\pm 0.57$ &  1&B220    & $ -1.21\pm 0.09$ &  1&B221    & $ -1.29\pm 0.04$ &  1&B222    & $ -0.93\pm 0.95$ &  1\\
B223    & $ -1.13\pm 0.51$ &  1&B224    & $ -1.80\pm 0.05$ &  1&B225    & $ -0.67\pm 0.12$ &  1&B230    & $ -2.17\pm 0.16$ &  1\\
B365    & $ -1.78\pm 0.19$ &  1&B231    & $ -1.49\pm 0.41$ &  1&B232    & $ -1.83\pm 0.14$ &  1&DAO070  & $  0.33\pm 0.36$ &  1\\
B281    & $ -0.87\pm 0.52$ &  1&B234    & $ -0.95\pm 0.13$ &  1&B366    & $ -1.79\pm 0.05$ &  1&B367    & $ -2.32\pm 0.53$ &  1\\
B283    & $ -0.06\pm 0.20$ &  1&B475    & $ -2.00\pm 0.14$ &  1&B235    & $ -0.72\pm 0.26$ &  1&DAO073  & $ -1.99\pm 0.19$ &  1\\
B237    & $ -2.09\pm 0.28$ &  1&B370    & $ -1.80\pm 0.02$ &  1&B238    & $ -0.57\pm 0.66$ &  1&B372    & $ -1.42\pm 0.17$ &  1\\
B374    & $ -1.90\pm 0.67$ &  1&B480    & $ -1.86\pm 0.66$ &  1&DAO084  & $ -1.79\pm 0.72$ &  1&B483    & $ -2.96\pm 0.35$ &  1\\
B484    & $ -1.95\pm 0.59$ &  1&B378    & $ -1.64\pm 0.26$ &  1&B380    & $ -2.31\pm 0.45$ &  1&B382    & $ -1.52\pm 0.27$ &  1\\
B386    & $ -1.62\pm 0.14$ &  1&B289D   & $ -1.71\pm 0.63$ &  1&B292D   & $ -0.47\pm 0.54$ &  1&G327    & $ -1.88\pm 0.06$ &  1\\
B391    & $ -0.55\pm 0.59$ &  1&B400    & $ -2.01\pm 0.21$ &  1&BA11    & $ -1.14\pm 0.61$ &  1&        &                  &   \\
\hline
\end{tabular}
\end{center}
{\sc Note} -- The spectroscopic metallicities used in this paper are
from \citet{per02} (source=1), \citet{bh00} (source=2) and
\citet{hbk91} (source=3).
\end{table*}

\section{Reddening Determinations}
\label{red.sec}

As already discussed in the introduction, there are several ways of
dealing with the problem of determining the reddening towards the M31
clusters. \citet{bh00} determined the largest number of reliable
reddening values for M31 GCs using correlations between optical and
infrared colours and metallicity, and by defining various
``reddening-free'' parameters based on their catalogue of multicolour
photometry. \citet{bh00} compared their results with those in the
literature and confirmed that their estimated reddening values are
reasonable, and quantitatively consistent with previous determinations
for GCs across the entire M31 disc. Below, we will determine more M31
GC reddening values based on the method of \citet{bh00}, and on the
Revised Bologna Catalogue \citep{gall04}, which is the newest and the
most comprehensive multicolour catalogue available at present.

\subsection{Constructing the correlations based on Galactic clusters}

In this section, we will construct the correlations between optical
colours and metallicity by defining various ``reddening-free''
parameters (henceforth referred to as $Q$ parameters), following
\citet{bh00}, based on the infrared photometry of \citet{bh90} and on
the updated Galactic GC catalogue of \citet[][updated in February
2003; hereafter H03]{harris96}. \citet{bh90} presented infrared
photometry for 23 Galactic GCs. H03 lists the reddening values,
metallicities, and optical colours of 150 Galactic GCs.

First, we performed linear regressions of intrinsic optical colours
versus metallicity. We use the Galactic GCs with $E(B-V)<0.5$ mag,
following \citet{bh00}:
\begin{equation}
(X-Y)_0=a\rm{[Fe/H]}+b \quad ;
\end{equation}

\begin{equation}
E(B-V)={E(B-V)\over E(X-Y)}[(X-Y)-(X-Y)_0] \quad ,
\end{equation}
where $(X-Y)$ represents any colour, and $(X-Y)_0$ represents the
relevant intrinsic colour obtained based on the reddening values
listed in H03. The reddening ratio can be determined from the Galactic
extinction law of \citet{car89}. The fit results with correlation
coefficients $r>0.8$ are listed in Table \ref{t3.tab}. We use
bi-sector linear fits, as described by \citet{Akritas96}, because we
are not only interested in the case where metallicity is used to
predict colour, but also in the reverse case where colour is used to
predict metallicity \citep[see details in][]{bh00}.

\begin{table*}
\caption{Colour-metallicity relations for Galactic GCs}
\label{t3.tab}
\begin{center}
\begin{tabular}{ccccccc}
\hline
\hline  & & \multicolumn{2}{c}{$(X-Y)_0=a{\rm [Fe/H]}+b$}& \multicolumn{2}{c}{${\rm [Fe/H]}= a(X-Y)_0+b$}  \\
\hline  $(X-Y)_0$& $N$ & $a$ & $b$ &  $a$ & $b$ & $r$ \\
\hline
$(U-B)_0$ & 84 &  0.316$\pm$ 0.000 & 0.599 $\pm$ 0.001 & 3.162 $\pm$ 0.028 &$ -1.895 \pm 0.002$ & 0.867 \\
$(U-V)_0$ & 84 &  0.491$\pm$ 0.001 & 1.538 $\pm$ 0.001 & 2.039 $\pm$ 0.010 &$ -3.136 \pm 0.008$ & 0.894 \\
$(U-R)_0$ & 67 & 0.559 $\pm$ 0.001 & 2.038 $\pm$ 0.002 & 1.788 $\pm$ 0.008 &$ -3.644 \pm 0.014$ & 0.906 \\
$(U-I)_0$ & 76 & 0.604 $\pm$ 0.002 & 2.555 $\pm$ 0.004 & 1.656 $\pm$ 0.012 &$ -4.232 \pm 0.004$ & 0.880 \\
$(U-J)_0$ & 32 & 0.966 $\pm$ 0.006 & 3.806 $\pm$ 0.011 & 1.035 $\pm$ 0.007 &$ -3.940 \pm 0.046$ & 0.883 \\
$(U-H)_0$ & 32 & 1.095 $\pm$ 0.007 & 4.561 $\pm$ 0.014 & 0.913 $\pm$ 0.005 &$ -4.166 \pm 0.051$ & 0.899 \\
$(U-K)_0$ & 32 & 1.151 $\pm$ 0.008 & 4.735 $\pm$ 0.015 & 0.868 $\pm$ 0.005 &$ -4.112 \pm 0.050$ & 0.897 \\
$(B-R)_0$ & 67 & 0.265 $\pm$ 0.000 & 1.474 $\pm$ 0.001 & 3.768 $\pm$ 0.070 &$ -5.553 \pm 0.088$ & 0.834 \\
$(B-J)_0$ & 32 & 0.694 $\pm$ 0.003 & 3.285 $\pm$ 0.005 & 1.440 $\pm$ 0.013 &$ -4.731 \pm 0.082$ & 0.848 \\
$(B-H)_0$ & 32 & 0.820 $\pm$ 0.004 & 4.034 $\pm$ 0.007 & 1.220 $\pm$ 0.009 &$ -4.922 \pm 0.083$ & 0.878 \\
$(B-K)_0$ & 32 & 0.877 $\pm$ 0.005 & 4.209 $\pm$ 0.007 & 1.140 $\pm$ 0.008 &$ -4.800 \pm 0.079$ & 0.875 \\
$(V-J)_0$ & 32 & 0.539 $\pm$ 0.002 & 2.378 $\pm$ 0.004 & 1.855 $\pm$ 0.027 &$ -4.413 \pm 0.084$ & 0.802 \\
$(V-H)_0$ & 32 & 0.660 $\pm$ 0.003 & 3.122 $\pm$ 0.005 & 1.514 $\pm$ 0.016 &$ -4.728 \pm 0.085$ & 0.850 \\
$(V-K)_0$ & 32 & 0.718 $\pm$ 0.003 & 3.298 $\pm$ 0.005 & 1.393 $\pm$ 0.013 &$ -4.594 \pm 0.078$ & 0.849 \\
$(R-H)_0$ & 24 & 0.694 $\pm$ 0.005 & 2.730 $\pm$ 0.008 & 1.442 $\pm$ 0.022 &$ -3.937 \pm 0.080$ & 0.852 \\
$(R-K)_0$ & 24 & 0.756 $\pm$ 0.006 & 2.912 $\pm$ 0.009 & 1.323 $\pm$ 0.017 &$ -3.851 \pm 0.070$ & 0.853 \\
\hline
\end{tabular}
\end{center}
\end{table*}

Next, we construct relationships between $Q$ parameters and intrinsic
colours, to estimate the reddening for clusters without spectroscopic
metallicities. The $Q$ parameters are defined as

\hspace{-12cm}{$$Q_{XYZ}\equiv(X-Y)-{E(X-Y)\over E(Y-Z)}(Y-Z)$$}
\begin{equation}
~~~~~~~~~~~~~~~~~~~~~~~~~~~~~~~~~=(X-Y)_0-{E(X-Y)\over
E(Y-Z)}(Y-Z)_0 \quad ,
\end{equation}
where $X, Y $ and $Z$ refer to photometric magnitudes in any
filter. The relation between an intrinsic colour and the $Q$ parameter
is
\begin{equation}
(X-Y)_0=a Q_{XYZ} + b \quad ,
\end{equation}

\begin{equation}
(X-Z)_0=a Q_{XYZ} + b \quad ,
\end{equation}
and
\begin{equation}
(Y-Z)_0=a Q_{XYZ} + b
\end{equation}

The fit results with correlation coefficients $r>0.8$ are listed in
Table \ref{t4.tab}.

\begin{table*}
\begin{center}
\caption{$Q$-parameter--colour relations for Galactic GCs}
\label{t4.tab}
\begin{tabular}{cccrcr}
\hline
\hline  $Q_{XYZ}$ & $(X-Y)_0$ &$N$ & \multicolumn{1}{c}{$a$} &  $b$  & \multicolumn{1}{c}{$r$}\\
\hline
$Q_{UBV}$& $U-B$ & 84 &$ 1.617 \pm 0.010$ & 0.759 $\pm$ 0.002 &$ 0.889$ \\
$Q_{UBR}$& $U-B$ & 67 &$ 1.423 \pm 0.002$ & 0.611 $\pm$ 0.000 &$ 0.966$ \\
$Q_{UBJ}$& $U-B$ & 32 &$ 1.740 \pm 0.032$ & 0.838 $\pm$ 0.005 &$ 0.888$ \\
$Q_{UBH}$& $U-B$ & 32 &$ 1.902 \pm 0.057$ & 1.058 $\pm$ 0.014 &$ 0.862$ \\
$Q_{UBK}$& $U-B$ & 32 &$ 1.907 \pm 0.058$ & 1.022 $\pm$ 0.013 &$ 0.861$ \\
$Q_{UVR}$& $U-V$ & 67 &$ 1.348 \pm 0.006$ & 0.987 $\pm$ 0.000 &$ 0.903$ \\
$Q_{VJK}$& $J-K$ & 32 &$-0.445 \pm 0.002$ & 0.096 $\pm$ 0.004 &$-0.842$ \\
$Q_{UBR}$& $U-R$ & 67 &$ 2.764 \pm 0.043$ & 2.144 $\pm$ 0.005 &$ 0.869$ \\
$Q_{UVR}$& $U-R$ & 67 &$ 1.637 \pm 0.019$ & 1.434 $\pm$ 0.001 &$ 0.835$ \\
$Q_{VRJ}$& $V-J$ & 24 &$-2.639 \pm 0.123$ & 0.988 $\pm$ 0.008 &$-0.842$ \\
$Q_{VRH}$& $V-H$ & 24 &$-3.088 \pm 0.128$ & 0.998 $\pm$ 0.021 &$-0.873$ \\
$Q_{RIH}$& $R-H$ & 24 &$-1.444 \pm 0.011$ & 0.556 $\pm$ 0.010 &$-0.947$ \\
$Q_{RIK}$& $R-K$ & 24 &$-1.880 \pm 0.021$ & 0.655 $\pm$ 0.011 &$-0.944$ \\
\hline
\end{tabular}
\end{center}
\end{table*}

\begin{figure*}
\resizebox{\hsize}{!}{\rotatebox{-90}{\includegraphics{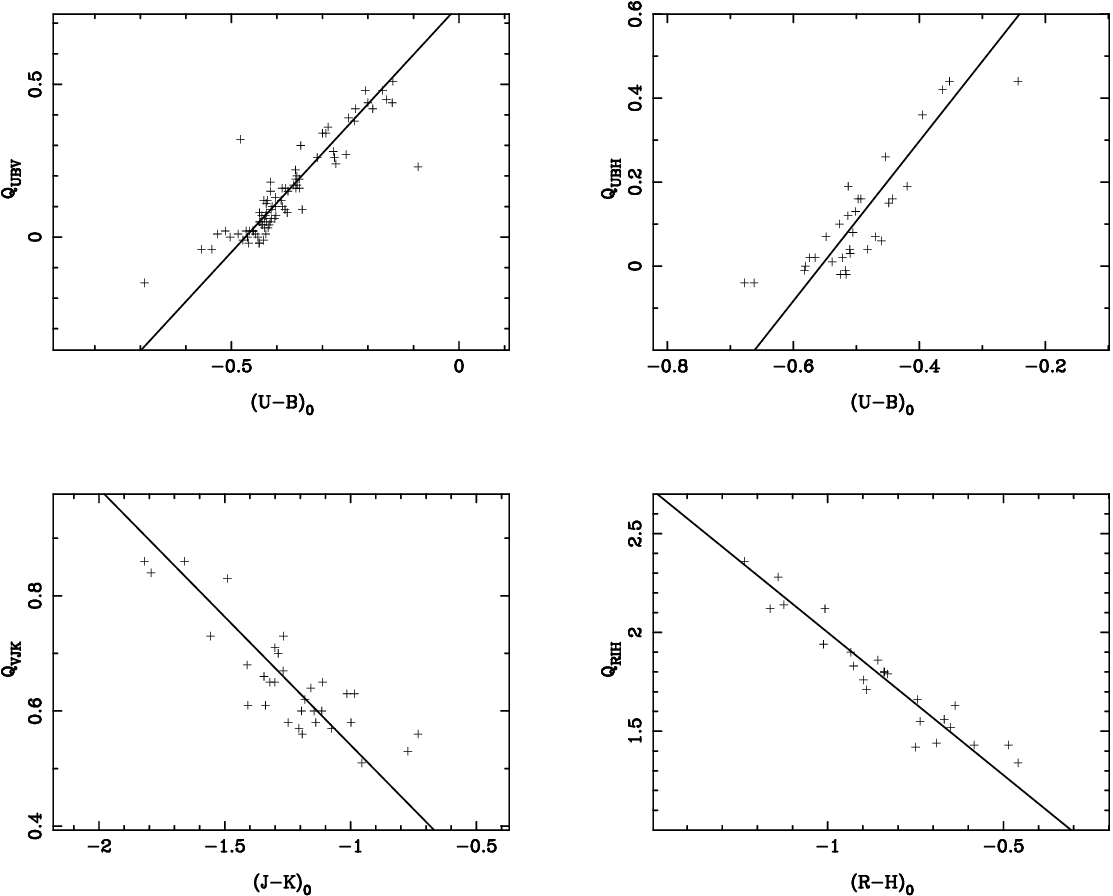}}}
\caption{Relationships between $Q$-parameter and colour for Galactic
GCs, for a few randomly selected representative relations from Table
3. We are unable to include error bars, since H03 does not provide
uncertainty estimates of their photometry.} \label{fig1}
\end{figure*}

Fig. 1 shows the relationships of a few representative fits between
$Q$-parameter and colour for Galactic GCs, randomly selected from
the set of relations included in Table 3.

\subsection{Tests of the methods using heavily reddened Galactic clusters}

In this section, we will test the methods adopted to derive the
reddening values, using the heavily reddened Galactic GCs with
$E(B-V)>0.5$ mag from H03 (which were not used to construct the
calibrations discussed above).

Based on Eqs. (1)--(6) and the correlation parameters from Tables
\ref{t3.tab} and \ref{t4.tab}, we can determine the reddening values
for these highly reddened Galactic GCs.

\begin{figure}
\resizebox{\hsize}{!}{\rotatebox{-90}{\includegraphics{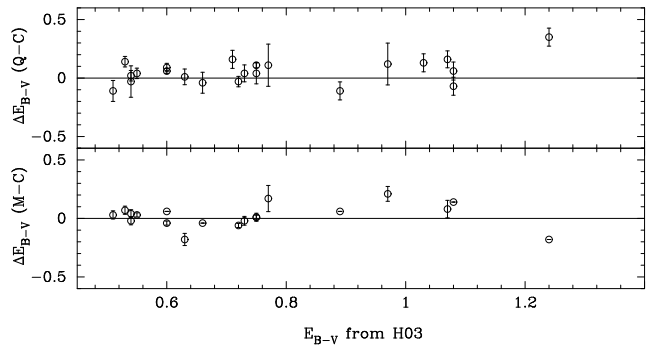}}}
\caption{Comparison of the reddening values between H03 and this
paper, for Galactic GCs. The error bars represent the uncertainties on
the reddening values derived in this paper only; H03 does not provide
uncertainty estimates.} \label{fig2}
\end{figure}

For each of the two methods we averaged all values of $E(B-V)$ to
produce one final value of $E(B-V)$ per method. The standard deviation
of the average value of $E(B-V)$ is taken as its error for each
method. The result of the comparison is shown in Fig. \ref{fig2},
from which we can see that the results are encouraging. The average
offset between $E(B-V)$ from the $Q$-parameter method and the H03
value is $0.01\pm0.01$ mag; for the colour-metallicity method, the
average offset is $0.00\pm0.01$ mag. It is clear that the two data
sets agree very well.

\subsection{Reddening values of the M31 clusters}

\citet{bh00} showed that the M31 and the Milky Way reddening laws are
the same within the observational errors. Therefore, in this section,
we will determine the reddening values for the M31 clusters and
cluster candidates based on the calibrated colour--metallicity (C-M)
and $Q$-parameter relations for the Milky Way from Tables \ref{t3.tab}
and \ref{t4.tab}. The metallicities are from the SMCat and the optical
and infrared photometric data are from \citet{gall04}, as discussed in
Sections 2.2 and 2.3. For each object, we average all reddening values
obtained using the various C-M and $Q$-colour relations, to get one
value for the reddening. The uncertainty in the reddening value thus
derived is calculated as the standard deviation of the resulting
reddening values.

We determined the reddening values for all M31 GCs and GC candidates
with sufficient data, a total of 658 objects. However, some
reddening values are not reliable, such as those based on only one
C-M or $Q$-colour relation, and those with large reddening errors.
In order to maintain consistency with \citet{bh00}, we adopted the
rules followed by these authors who rejected reddening values
$\sigma_{E(B-V)}~/~ \overline{E(B-V)}>0.5$ for
$\overline{E(B-V)}>0.15$ mag and $\sigma_{E(B-V)}~/~
\overline{E(B-V)}>1.0$ for $\overline{E(B-V)}<0.15$ mag. However,
following \citet{bh00}, we emphasize that the rules adopted here for
rejecting reddening values are quite arbitrary. In total, 443
reliable reddening values were determined in this paper, which are
listed in Table \ref{t5.tab}. Columns 1, 3, 5, 7, and 9 list the
names of the GCs, using the nomenclature adopted by \citet{gall04}.

From Sections 3.1 and 3.2 we find that the reddening values for the
Galactic GCs obtained from different relations are internally
consistent. However, for M31 GCs this is not always the case. For some
GCs and GC candidates the reddening values, based on different
relations, are inconsistent. Reasons for this may include that for
Galactic GCs, the photometric data are accurate, but for some M31 GCs
and GC candidates (particularly the fainter objects) this may not be
the case. Therefore, when we average the reddening values for each
object, we reject reddening values that are clearly statistical
outliers: these are defined as those values that differ from the mean
value for a given object by more than $1\sigma$.

\begin{table*}
\caption{Reliable reddening values of GCs and GC candidates in
M31.} \label{t5.tab}
\begin{center}
\begin{tabular}{lclclclclc}
\hline
\hline \multicolumn{1}{c}{Name}& $E(B-V)$ &\multicolumn{1}{c}{Name}& $E(B-V)$& \multicolumn{1}{c}{Name}& $E(B-V)$ & \multicolumn{1}{c}{Name}& $E(B-V)$ & \multicolumn{1}{c}{Name}& $E(B-V)$ \\
\hline

G001    & $  0.10\pm 0.02$ &G002    & $  0.05\pm 0.01$ &B290    & $  0.13\pm 0.03$ &BA21    & $  0.35\pm 0.07$ &B412    & $  0.26\pm 0.02$ \\
B413    & $  0.48\pm 0.04$ &B134D   & $  0.45\pm 0.09$ &B291    & $  0.05\pm 0.02$ &B138D   & $  0.23\pm 0.04$ &B140D   & $  0.45\pm 0.11$ \\
B141D   & $  0.43\pm 0.09$ &B142D   & $  0.58\pm 0.03$ &B144D   & $  0.33\pm 0.02$ &B147D   & $  0.33\pm 0.05$ &B295    & $  0.10\pm 0.01$ \\
B148D   & $  0.04\pm 0.03$ &B149D   & $  0.69\pm 0.00$ &B150D   & $  0.32\pm 0.02$ &B416    & $  0.54\pm 0.05$ &B152D   & $  0.52\pm 0.19$ \\
B418    & $  0.26\pm 0.09$ &B154D   & $  0.09\pm 0.07$ &B156D   & $  0.45\pm 0.09$ &B420    & $  0.30\pm 0.03$ &B157D   & $  0.15\pm 0.01$ \\
B422    & $  0.10\pm 0.08$ &B162D   & $  0.53\pm 0.07$ &B163D   & $  0.16\pm 0.03$ &B298    & $  0.16\pm 0.02$ &B424    & $  0.59\pm 0.03$ \\
B165D   & $  0.21\pm 0.07$ &B166D   & $  0.26\pm 0.04$ &B301    & $  0.17\pm 0.02$ &B167D   & $  0.23\pm 0.03$ &B427    & $  0.53\pm 0.03$ \\
B169D   & $  0.07\pm 0.03$ &B170D   & $  0.23\pm 0.11$ &B302    & $  0.10\pm 0.01$ &B428    & $  0.54\pm 0.05$ &B172D   & $  0.11\pm 0.05$ \\
B430    & $  0.10\pm 0.05$ &B173D   & $  0.34\pm 0.03$ &B175D   & $  0.08\pm 0.03$ &B303    & $  0.14\pm 0.06$ &B177D   & $  0.08\pm 0.02$ \\
B304    & $  0.07\pm 0.01$ &B433    & $  0.61\pm 0.05$ &B306    & $  0.42\pm 0.02$ &B435    & $  0.67\pm 0.05$ &B307    & $  0.08\pm 0.02$ \\
B178D   & $  0.14\pm 0.02$ &B309    & $  0.17\pm 0.04$ &B310    & $  0.09\pm 0.01$ &B181D   & $  0.36\pm 0.09$ &B311    & $  0.29\pm 0.02$ \\
B438    & $  0.82\pm 0.07$ &B312    & $  0.16\pm 0.01$ &B183D   & $  0.36\pm 0.05$ &B314    & $  0.08\pm 0.05$ &B313    & $  0.21\pm 0.02$ \\
B315    & $  0.07\pm 0.02$ &B001    & $  0.25\pm 0.02$ &B316    & $  0.21\pm 0.03$ &B317    & $  0.11\pm 0.02$ &B186D   & $  0.33\pm 0.05$ \\
B440    & $  0.32\pm 0.15$ &B003    & $  0.19\pm 0.02$ &B188D   & $  0.56\pm 0.06$ &B190D   & $  0.26\pm 0.03$ &B004    & $  0.07\pm 0.02$ \\
B005    & $  0.28\pm 0.02$ &B325    & $  0.14\pm 0.02$ &B328    & $  0.10\pm 0.02$ &B192D   & $  0.45\pm 0.02$ &B330    & $  0.31\pm 0.03$ \\
B004D   & $  0.57\pm 0.07$ &B006    & $  0.09\pm 0.02$ &B194D   & $  0.53\pm 0.04$ &B447    & $  0.34\pm 0.13$ &B244    & $  0.27\pm 0.03$ \\
B009    & $  0.13\pm 0.02$ &B010    & $  0.22\pm 0.01$ &B011    & $  0.11\pm 0.01$ &B012    & $  0.12\pm 0.01$ &B196D   & $  0.19\pm 0.06$ \\
B245    & $  1.37\pm 0.07$ &B448    & $  0.05\pm 0.01$ &B013    & $  0.13\pm 0.02$ &B014    & $  0.36\pm 0.02$ &B197D   & $  0.39\pm 0.05$ \\
B335    & $  0.65\pm 0.02$ &B449    & $  1.27\pm 0.08$ &B015    & $  0.50\pm 0.02$ &B016    & $  0.30\pm 0.02$ &B450    & $  0.24\pm 0.10$ \\
B337    & $  0.06\pm 0.02$ &B017    & $  0.27\pm 0.02$ &B018    & $  0.20\pm 0.01$ &B019    & $  0.20\pm 0.01$ &B020    & $  0.12\pm 0.01$ \\
B338    & $  0.14\pm 0.02$ &B021    & $  0.26\pm 0.02$ &B022    & $  0.04\pm 0.03$ &B339    & $  0.16\pm 0.03$ &B023    & $  0.32\pm 0.01$ \\
B453    & $  0.30\pm 0.02$ &B024    & $  0.03\pm 0.02$ &V031    & $  0.33\pm 0.02$ &B025    & $  0.20\pm 0.01$ &B202D   & $  0.38\pm 0.07$ \\
B027    & $  0.21\pm 0.01$ &B026    & $  0.15\pm 0.02$ &B028    & $  0.22\pm 0.02$ &B020D   & $  0.22\pm 0.06$ &B029    & $  0.12\pm 0.01$ \\
B030    & $  0.48\pm 0.03$ &B031    & $  0.33\pm 0.02$ &B032    & $  0.42\pm 0.02$ &B456    & $  0.32\pm 0.04$ &B203D   & $  0.36\pm 0.04$ \\
B033    & $  0.14\pm 0.02$ &B034    & $  0.19\pm 0.01$ &B457    & $  0.14\pm 0.02$ &B036    & $  0.15\pm 0.02$ &B204D   & $  0.48\pm 0.22$ \\
B025D   & $  0.58\pm 0.03$ &B037    & $  1.21\pm 0.03$ &B038    & $  0.27\pm 0.01$ &B039    & $  0.38\pm 0.02$ &B205D   & $  0.78\pm 0.06$ \\
B041    & $  0.07\pm 0.03$ &B042    & $  0.61\pm 0.01$ &B044    & $  0.33\pm 0.01$ &B343    & $  0.06\pm 0.01$ &B045    & $  0.18\pm 0.01$ \\
B046    & $  0.19\pm 0.03$ &B207D   & $  0.33\pm 0.08$ &B048    & $  0.19\pm 0.02$ &B047    & $  0.09\pm 0.02$ &B049    & $  0.16\pm 0.02$ \\
B050    & $  0.24\pm 0.01$ &B051    & $  0.34\pm 0.02$ &B052    & $  0.23\pm 0.04$ &B054    & $  0.23\pm 0.02$ &B056    & $  0.17\pm 0.01$ \\
B057    & $  0.09\pm 0.02$ &B058    & $  0.13\pm 0.01$ &B059    & $  0.29\pm 0.01$ &B060    & $  0.21\pm 0.02$ &B061    & $  0.34\pm 0.02$ \\
B062    & $  0.26\pm 0.03$ &B063    & $  0.40\pm 0.01$ &B064    & $  0.17\pm 0.01$ &B065    & $  0.10\pm 0.01$ &B344    & $  0.11\pm 0.02$ \\
B067    & $  0.24\pm 0.03$ &B068    & $  0.38\pm 0.03$ &B257    & $  1.17\pm 0.03$ &B461    & $  0.58\pm 0.07$ &B070    & $  0.12\pm 0.04$ \\
B073    & $  0.11\pm 0.01$ &B072    & $  0.33\pm 0.06$ &B074    & $  0.19\pm 0.01$ &B075    & $  0.30\pm 0.04$ &B077    & $  0.97\pm 0.03$ \\
B078    & $  0.44\pm 0.07$ &B079    & $  0.58\pm 0.03$ &B081    & $  0.11\pm 0.02$ &B345    & $  0.10\pm 0.02$ &B462    & $  0.39\pm 0.04$ \\
B082    & $  0.62\pm 0.03$ &B083    & $  0.12\pm 0.02$ &B084    & $  0.26\pm 0.04$ &B085    & $  0.14\pm 0.02$ &B086    & $  0.15\pm 0.01$ \\
B346    & $  0.71\pm 0.03$ &B088    & $  0.46\pm 0.01$ &B092    & $  0.12\pm 0.02$ &B347    & $  0.14\pm 0.02$ &B348    & $  0.25\pm 0.04$ \\
B093    & $  0.30\pm 0.02$ &B094    & $  0.07\pm 0.02$ &B095    & $  0.43\pm 0.04$ &B096    & $  0.26\pm 0.02$ &B098    & $  0.08\pm 0.02$ \\
B463    & $  0.33\pm 0.07$ &B097    & $  0.29\pm 0.01$ &B099    & $  0.16\pm 0.03$ &B350    & $  0.10\pm 0.02$ &B100    & $  0.48\pm 0.08$ \\
B101    & $  0.17\pm 0.02$ &NB46    & $  0.62\pm 0.07$ &B103    & $  0.19\pm 0.02$ &NB70    & $  0.39\pm 0.03$ &B464    & $  0.27\pm 0.03$ \\
B105    & $  0.14\pm 0.01$ &B106    & $  0.12\pm 0.02$ &B107    & $  0.28\pm 0.02$ &B109    & $  0.08\pm 0.02$ &B111    & $  0.08\pm 0.02$ \\
B110    & $  0.20\pm 0.02$ &B260    & $  0.67\pm 0.02$ &B112    & $  0.14\pm 0.02$ &B117    & $  0.04\pm 0.01$ &B115    & $  0.12\pm 0.01$ \\
B116    & $  0.62\pm 0.02$ &NB64    & $  0.46\pm 0.15$ &B118    & $  0.22\pm 0.06$ &B119    & $  0.14\pm 0.02$ &B351    & $  0.15\pm 0.02$ \\
B352    & $  0.14\pm 0.02$ &NB25    & $  0.66\pm 0.17$ &B122    & $  0.80\pm 0.02$ &B123    & $  0.30\pm 0.03$ &B125    & $  0.05\pm 0.02$ \\
B127    & $  0.09\pm 0.02$ &B354    & $  0.05\pm 0.02$ &B128    & $  0.15\pm 0.01$ &B129    & $  1.16\pm 0.06$ &NB39    & $  0.48\pm 0.02$ \\
B130    & $  0.36\pm 0.01$ &B131    & $  0.12\pm 0.04$ &B134    & $  0.03\pm 0.02$ &B135    & $  0.27\pm 0.01$ &B355    & $  0.07\pm 0.06$ \\
B136    & $  0.36\pm 0.04$ &B137    & $  0.40\pm 0.02$ &B217D   & $  0.18\pm 0.01$ &B141    & $  0.32\pm 0.01$ &B143    & $  0.05\pm 0.02$ \\
B144    & $  0.05\pm 0.02$ &B219D   & $  0.42\pm 0.02$ &B146    & $  0.06\pm 0.04$ &B266    & $  0.98\pm 0.09$ &B148    & $  0.17\pm 0.02$ \\
B220D   & $  0.07\pm 0.04$ &B149    & $  0.34\pm 0.02$ &B221D   & $  0.53\pm 0.08$ &B467    & $  0.27\pm 0.02$ &B150    & $  0.28\pm 0.07$ \\
B223D   & $  0.20\pm 0.05$ &B151    & $  0.32\pm 0.01$ &B152    & $  0.18\pm 0.01$ &B356    & $  0.31\pm 0.01$ &B153    & $  0.05\pm 0.01$ \\
B154    & $  0.15\pm 0.03$ &B468    & $  0.27\pm 0.03$ &B357    & $  0.12\pm 0.02$ &B155    & $  0.20\pm 0.02$ &B156    & $  0.10\pm 0.02$ \\
B158    & $  0.14\pm 0.00$ &B159    & $  0.36\pm 0.04$ &B226D   & $  0.63\pm 0.00$ &B161    & $  0.17\pm 0.01$ &B162    & $  0.25\pm 0.03$ \\
B163    & $  0.14\pm 0.01$ &B358    & $  0.06\pm 0.01$ &B164    & $  0.12\pm 0.03$ &B165    & $  0.12\pm 0.01$ &B228D   & $  0.18\pm 0.02$ \\
B167    & $  0.03\pm 0.02$ &B168    & $  0.54\pm 0.05$ &B169    & $  0.59\pm 0.04$ &B170    & $  0.10\pm 0.02$ &B272    & $  0.57\pm 0.04$ \\
B171    & $  0.11\pm 0.01$ &B172    & $  0.18\pm 0.02$ &DAO062  & $  1.11\pm 0.17$ &B173    & $  0.40\pm 0.04$ &B174    & $  0.32\pm 0.01$ \\
B177    & $  0.18\pm 0.03$ &B176    & $  0.10\pm 0.01$ &B178    & $  0.12\pm 0.01$ &B179    & $  0.10\pm 0.01$ &B180    & $  0.14\pm 0.01$ \\
B181    & $  0.23\pm 0.01$ &B231D   & $  0.07\pm 0.02$ &B182    & $  0.25\pm 0.01$ &B185    & $  0.11\pm 0.01$ &B184    & $  0.21\pm 0.03$ \\
B470    & $  0.40\pm 0.08$ &B187    & $  0.35\pm 0.02$ &B471    & $  0.62\pm 0.04$ &B189    & $  0.04\pm 0.03$ &B190    & $  0.10\pm 0.02$ \\
B192    & $  0.31\pm 0.02$ &B194    & $  0.07\pm 0.02$ &B193    & $  0.11\pm 0.01$ &B472    & $  0.13\pm 0.00$ &B195    & $  0.12\pm 0.00$ \\
B196    & $  0.26\pm 0.04$ &B235D   & $  0.49\pm 0.03$ &B197    & $  0.19\pm 0.02$ &B199    & $  0.10\pm 0.02$ &B201    & $  0.04\pm 0.02$ \\
B202    & $  0.26\pm 0.02$ &B203    & $  0.16\pm 0.02$ &B204    & $  0.12\pm 0.01$ &B361    & $  0.11\pm 0.01$ &B237D   & $  0.32\pm 0.05$ \\
B205    & $  0.14\pm 0.01$ &B206    & $  0.13\pm 0.01$ &B238D   & $  0.40\pm 0.07$ &B208    & $  0.13\pm 0.02$ &G260    & $  0.30\pm 0.05$ \\
\hline
\end{tabular}
\end{center}
\end{table*}
\addtocounter{table}{-1}

\begin{table*}
\caption{Continued.}
\label{t6.tab}
\begin{center}
\begin{tabular}{lclclclclc}
\hline
\hline \multicolumn{1}{c}{Name}& $E(B-V)$ &\multicolumn{1}{c}{Name}& $E(B-V)$& \multicolumn{1}{c}{Name}& $E(B-V)$ & \multicolumn{1}{c}{Name}& $E(B-V)$ & \multicolumn{1}{c}{Name}& $E(B-V)$ \\
\hline

B239D   & $  0.38\pm 0.12$ &B209    & $  0.10\pm 0.01$ &B211    & $  0.10\pm 0.01$ &B212    & $  0.13\pm 0.01$ &B213    & $  0.15\pm 0.02$ \\
B214    & $  0.05\pm 0.02$ &B215    & $  0.21\pm 0.04$ &B362    & $  0.65\pm 0.20$ &G268    & $  0.26\pm 0.01$ &B217    & $  0.12\pm 0.01$ \\
G270    & $  0.67\pm 0.20$ &B218    & $  0.14\pm 0.01$ &B219    & $  0.05\pm 0.03$ &B243D   & $  0.03\pm 0.02$ &B220    & $  0.05\pm 0.02$ \\
B245D   & $  0.52\pm 0.03$ &B221    & $  0.23\pm 0.02$ &B222    & $  0.05\pm 0.02$ &B224    & $  0.13\pm 0.02$ &B473    & $  0.20\pm 0.04$ \\
B225    & $  0.10\pm 0.01$ &B226    & $  1.08\pm 0.06$ &B247D   & $  0.51\pm 0.08$ &B227    & $  0.37\pm 0.04$ &B228    & $  0.13\pm 0.01$ \\
B229    & $  0.07\pm 0.02$ &B230    & $  0.15\pm 0.01$ &B365    & $  0.19\pm 0.02$ &B231    & $  0.15\pm 0.03$ &B232    & $  0.14\pm 0.01$ \\
B233    & $  0.17\pm 0.01$ &B281    & $  0.12\pm 0.02$ &B250D   & $  0.54\pm 0.06$ &B252D   & $  0.26\pm 0.04$ &B234    & $  0.11\pm 0.02$ \\
B366    & $  0.05\pm 0.02$ &B474    & $  0.53\pm 0.05$ &B475    & $  0.16\pm 0.03$ &B235    & $  0.11\pm 0.01$ &B256D   & $  0.69\pm 0.06$ \\
B476    & $  0.08\pm 0.05$ &B236    & $  0.07\pm 0.05$ &B258D   & $  1.20\pm 0.09$ &B237    & $  0.14\pm 0.02$ &B260D   & $  0.20\pm 0.03$ \\
B478    & $  1.00\pm 0.13$ &B370    & $  0.34\pm 0.01$ &B238    & $  0.11\pm 0.02$ &B239    & $  0.09\pm 0.01$ &B261D   & $  0.27\pm 0.06$ \\
B263D   & $  0.25\pm 0.07$ &B240    & $  0.13\pm 0.00$ &B286    & $  0.67\pm 0.02$ &B479    & $  0.64\pm 0.11$ &B266D   & $  0.75\pm 0.15$ \\
B372    & $  0.20\pm 0.02$ &B373    & $  0.10\pm 0.01$ &B375    & $  0.29\pm 0.03$ &B481    & $  0.52\pm 0.08$ &B377    & $  0.16\pm 0.02$ \\
B270D   & $  0.25\pm 0.02$ &B483    & $  0.08\pm 0.06$ &B378    & $  0.14\pm 0.02$ &B379    & $  0.15\pm 0.01$ &B273D   & $  0.29\pm 0.04$ \\
B274D   & $  0.23\pm 0.04$ &B380    & $  0.06\pm 0.02$ &B381    & $  0.17\pm 0.02$ &B275D   & $  0.18\pm 0.03$ &B486    & $  0.17\pm 0.02$ \\
B277D   & $  0.34\pm 0.08$ &B382    & $  0.10\pm 0.02$ &B278D   & $  0.42\pm 0.04$ &B384    & $  0.04\pm 0.02$ &B385    & $  0.06\pm 0.05$ \\
B386    & $  0.21\pm 0.01$ &B283D   & $  0.16\pm 0.04$ &B288D   & $  0.58\pm 0.06$ &B387    & $  0.12\pm 0.02$ &B489    & $  0.17\pm 0.04$ \\
B289D   & $  0.23\pm 0.05$ &B490    & $  0.30\pm 0.02$ &G325    & $  0.12\pm 0.09$ &B389    & $  0.27\pm 0.08$ &B293D   & $  0.27\pm 0.06$ \\
G327    & $  0.18\pm 0.01$ &B295D   & $  0.62\pm 0.12$ &B296D   & $  0.06\pm 0.04$ &B297D   & $  0.30\pm 0.07$ &B298D   & $  0.70\pm 0.11$ \\
B299D   & $  0.39\pm 0.00$ &B300D   & $  0.74\pm 0.15$ &B393    & $  0.14\pm 0.02$ &B492    & $  0.26\pm 0.03$ &B302D   & $  0.25\pm 0.02$ \\
B304D   & $  0.67\pm 0.04$ &B493    & $  0.09\pm 0.07$ &B494    & $  0.57\pm 0.07$ &B307D   & $  0.45\pm 0.08$ &B308D   & $  0.12\pm 0.07$ \\
B495    & $  0.34\pm 0.08$ &B396    & $  0.09\pm 0.01$ &B310D   & $  0.34\pm 0.03$ &B313D   & $  0.11\pm 0.04$ &B314D   & $  0.12\pm 0.03$ \\
B317D   & $  0.69\pm 0.03$ &B319D   & $  0.44\pm 0.06$ &B320D   & $  0.50\pm 0.06$ &B324D   & $  0.11\pm 0.06$ &B398    & $  0.16\pm 0.03$ \\
B399    & $  0.03\pm 0.02$ &B400    & $  0.21\pm 0.02$ &B326D   & $  0.31\pm 0.12$ &B328D   & $  0.65\pm 0.09$ &B329D   & $  0.16\pm 0.03$ \\
B330D   & $  0.70\pm 0.02$ &B331D   & $  0.50\pm 0.06$ &B332D   & $  0.33\pm 0.13$ &B402    & $  0.16\pm 0.03$ &BA11    & $  0.06\pm 0.03$ \\
B334D   & $  0.42\pm 0.05$ &B338D   & $  0.55\pm 0.09$ &B339D   & $  0.61\pm 0.05$ &B403    & $  0.07\pm 0.02$ &B340D   & $  0.23\pm 0.06$ \\
B405    & $  0.14\pm 0.02$ &B508    & $  0.10\pm 0.08$ &B343D   & $  0.18\pm 0.03$ &B344D   & $  0.07\pm 0.04$ &B345D   & $  0.08\pm 0.05$ \\
B346D   & $  0.19\pm 0.06$ &B407    & $  0.16\pm 0.02$ &B347D   & $  0.16\pm 0.05$ &B348D   & $  0.42\pm 0.04$ &B349D   & $  0.21\pm 0.01$ \\
NB63    & $  0.90\pm 0.05$ &NB16    & $  0.41\pm 0.08$ &NB50    & $  0.93\pm 0.21$ &        &                  &        &                  \\
\hline
\end{tabular}
\end{center}
\end{table*}

\begin{figure}
\resizebox{\hsize}{!}{\rotatebox{-90}{\includegraphics{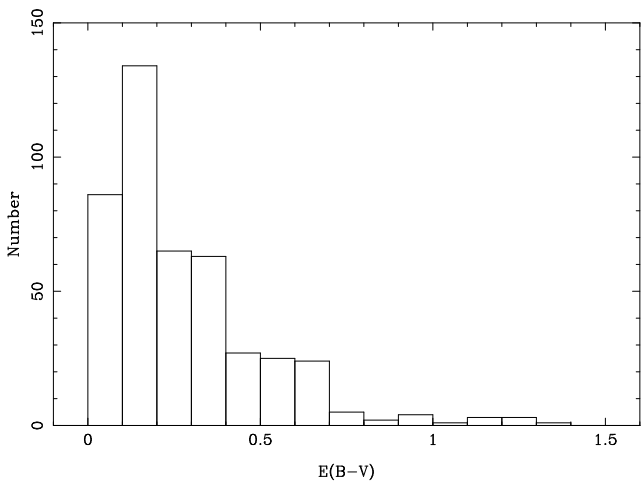}}}
\caption{Distribution of the reddening values of the M31 GCs and GC
candidates obtained in this paper.} \label{fig3}
\end{figure}

Fig. \ref{fig3} shows the distribution of the reliable reddening
values listed in Table \ref{t5.tab}. From Fig. \ref{fig3} we find that
slightly more than half of the reddening values are $E(B-V) < 0.2$
mag. The distribution of the 443 reliable reddening values has a mean
of $E(B-V) = 0.28_{-0.14}^{+0.23}$ with a standard deviation of
$\sigma=0.17$ mag, compared with $E(B-V)=0.22; \sigma=0.19$ mag of
\citet{bh00}.

\begin{figure*}
\resizebox{\hsize}{!}{\rotatebox{0}{\includegraphics{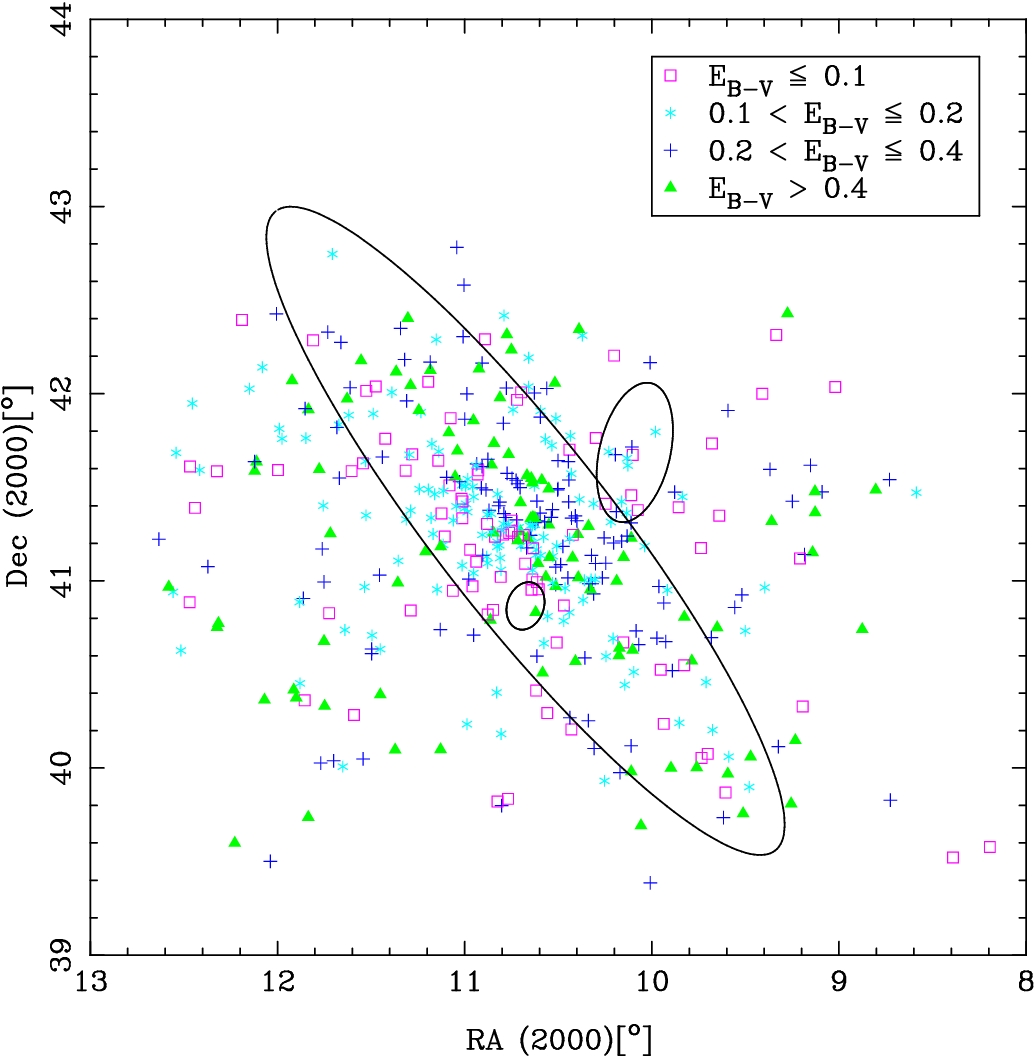}}}
\caption{Map of the confirmed and candidate GCs in M31. The large
ellipse is the M31 disc/halo boundary as defined by \citet{rac91};
the two small ellipses are the $D_{25}$ isophotes of NGC 205
(northwest) and M32 (southeast).} \label{fig4}
\end{figure*}

Fig. \ref{fig4} shows the reddening values as a function of
position. The large ellipse represents the boundary of the M31 disc
defined by \citet{rac91} and the small ellipses on the northwestern
and southeastern sides of the major axis are the $D_{25}$ diameters of
the M31 companion galaxies NGC 205 and M32, respectively. The
distribution appears reasonable in that the objects with low reddening
values are spread across the disc and halo, while those with high
reddening values are mainly concentrated in the galactic
disc. However, from Fig. \ref{fig4}, we can also see that a
substantial number of objects outside the ``halo'' boundary have
$E(B-V)>0.1$ mag, i.e., greater than the Galactic foreground reddening
in the direction of M31, as estimated by many authors \citep[see,
e.g.,][]{van69,McRa69,Frogel80}. In fact, \citet{bh00} also noted this
phenomenon. They suggested a number of plausible explanations, which
include that (i) this could be caused by the large uncertainties
inherent to the method; or (ii) the assumption that the M31 halo
clusters are subject to only foreground reddening is somewhat dubious.

\citet{ve62b} analysed the photometry of M31 GCs published by
\cite{ve62}. He assumed that the halo clusters are only affected by
foreground Galactic extinction. He derived a mean intrinsic colour of
$(B-V)_0=0.83$ mag from the objects in the halo of M31 and calculated
the colour excess for each object. Subsequently, he studied the
reddening distribution of M31 GCs on either side of the major axis and
found that the GCs on the northwestern side of the disc are either
intrinsically redder, or suffer from more significant extinction, than
those on the southeastern side of the disc.

Based on three homogeneous and independent photometric data sets for
M31 GC candidates, \citet{ir85} examined the differential reddening
towards these objects, and found that the GCs on the northwestern side
of the disc are redder than those on the southeastern side. Therefore,
\citet{ir85} concluded that the northwestern side of the disc of M31
is nearer to us than its southeastern side.

With the large sample of M31 clusters (and candidates) at hand, we can
now also examine the reddening distribution of M31 objects on either
side of the major axis, and calculate the mean $E(B-V)$.

\begin{figure}
\begin{center}
\resizebox{\hsize}{!}{\rotatebox{-90}{\includegraphics{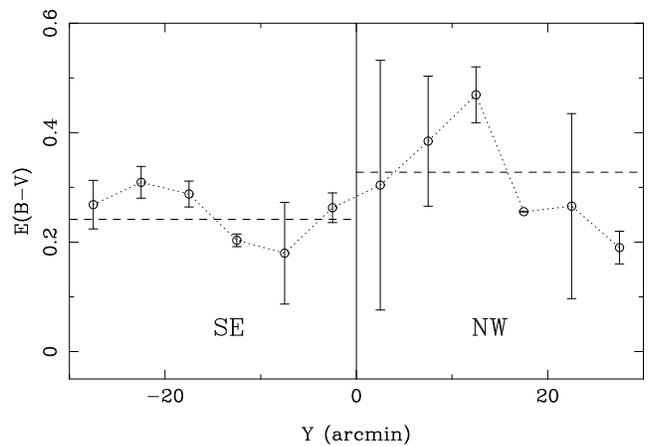}}}
\caption{Reddening distribution dependence on the distance from
the major axis of M31. The dashed lines indicate the mean
reddening values for GCs on the two sides.} \label{fig5}
\end{center}
\end{figure}

Following \citet{per02} and \citet{hbk91}, we use the $X,Y$ plane to
indicate the position of the GCs. The $X$ coordinate is the position
along the major axis of M31, where positive $X$ is in the northeastern
direction, while the $Y$ coordinate is along the minor axis of the M31
disc, increasing towards the northwest. The relative coordinates of
the M31 clusters are derived by assuming standard geometric parameters
for M31. We adopted a central position for M31 at $\rm
\alpha_0=00^h42^m44^s.30$ and $\rm \delta_0=+41^o16'09''.0$ (J2000.0)
following \citet{hbk91} and \citet{per02}. Formally,
\begin{equation}
X=A\sin\theta+B\cos\theta \quad ;
\end{equation}

\begin{equation}
Y=-A\cos\theta+B\sin\theta \quad ,
\end{equation}
where $A=\sin(\alpha-\alpha_0)\cos\delta$ and $B=\sin\delta
\cos\delta_0 - \cos(\alpha-\alpha_0) \cos\delta \sin\delta_0$. We
adopt a position angle of $\theta=38^\circ$ for the major axis of M31
\citep{ken89}. Fig. \ref{fig5} shows graphically the dependence of the
average reddening on the distance from the major axis. The error bars
represent the standard deviations of the means.  It is clear that the
reddening on the northwestern (NW) side of the disc is much greater
than that on the southeastern (SE) side. The mean reddening on the
northwestern and southeastern sides is $\overline{E(B-V)}=0.33 \pm
0.02$ and $0.24 \pm 0.02$ mag, respectively.

\begin{figure}
\begin{center}
\resizebox{\hsize}{!}{\rotatebox{-90}{\includegraphics{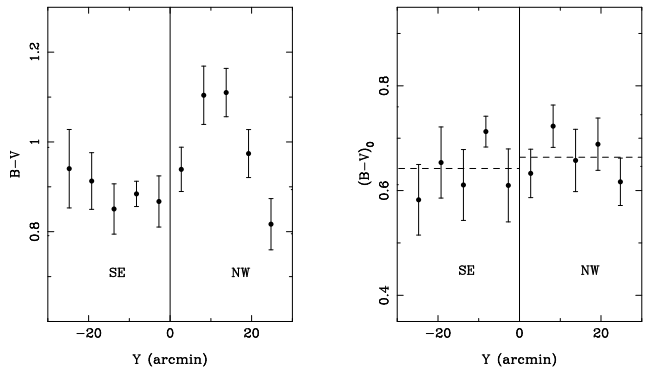}}}
\caption{$(B-V)$ colour and intrinsic $(B-V)_0$ colour versus distance
from the major axis of M31 for M31 GCs and GC candidates.}
\label{fig6}
\end{center}
\end{figure}

Below, we will check our resulting reddening values by studying the
distribution of the colours of the M31 clusters and cluster
candidates as a function of the projected distance, $Y$, from the
major axis. If our reddening values are correct, the distribution of
the {\it intrinsic} colours of the M31 clusters should be symmetric,
even if the distribution of the {\it observed} colours is
asymmetric. The left-hand panel of Fig. \ref{fig6} shows the
distribution of the mean colours of the GCs and GC candidates binned
in 5.5 arcmin intervals in $Y$. The error bars represent the
standard deviations of the means. It is clear that the colour
distribution is asymmetric. The right-hand panel of Fig. \ref{fig6}
shows the distribution of the mean intrinsic colours of the GCs and
GC candidates binned in 5.5 arcmin intervals in $Y$. Again, the
error bars represent the standard deviations of the means.
Obviously, the distribution of the mean intrinsic colours is nearly
symmetric. On the southeastern side, the mean intrinsic colour is
$\overline{(B-V)_0}$= 0.64$\pm$0.03, while on the northwestern side
the mean intrinsic colour is $\overline{(B-V)_0}$=0.66$\pm$0.02.

\section{M31 Globular Cluster Metallicities}
\label{metal.sec}

It is evident that the metallicity distribution of a galaxy's GC
system can provide important clues as to the process and conditions
relevant to galaxy formation. Previous studies of the M31 GC system
have presented some important information. For example, signs of
bimodality were found in the M31 metallicity distribution by
\citet{hbk91}, \citet{ash93}, \citet{bh00} and \citet{per02}. In
addition, \citet{hbk91} found that the metal-rich clusters in M31
appear to form a central rotating disc system. Based on the current
largest sample including 321 velocities, \citet{per02} performed a
more comprehensive investigation into the kinematics of the M31
cluster system and showed that the metal-rich GCs appear to constitute
a distinct kinematic subsystem that demonstrates a centrally
concentrated spatial distribution with a high rotation amplitude, but
it does not appear to be significantly flattened. This is consistent
with a bulge population. In this section, we will examine the
distribution of GC metallicities in M31 using the largest number of
GCs and GC candidates available to date. We will include the
metallicities determined based on GC colours.

\subsection{Colour-derived metallicities}

In total, 231 of the GCs and GC candidates in our sample of 443
objects with reliable reddening values have no spectroscopic
metallicities. Therefore, we will determine their metallicities based
on the C-M relation for these objects. From the colour excesses of GCs
with spectroscopic metallicities, \citet{bh00} found that the M31 and
Galactic GC C-M relations are consistent.  Therefore, in this section,
we will determine metallicities for 231 GCs and GC candidates from
their intrinsic colours by applying the Galactic C-M relation. We use
bi-sector linear fits \citep{Akritas96} to determine the metallicity
as a function of colour, and average the resulting metallicities over
the available colours for each object (except for some significantly
deviating points that are most likely due to inaccurate photometric
data, i.e.  those that differ from the mean value by more than
$1\sigma$, as justified above). As in the reddening determination, the
standard deviations of the metallicities from individual colours are
used as the error estimates. There are 209 GCs (and candidates) with
metallicity determinations, which are listed in Table \ref{t7.tab}.

\subsection{Comparison of spectroscopic and colour-derived metallicities}

\begin{figure}
\begin{center}
\resizebox{\hsize}{!}{\rotatebox{0}{\includegraphics{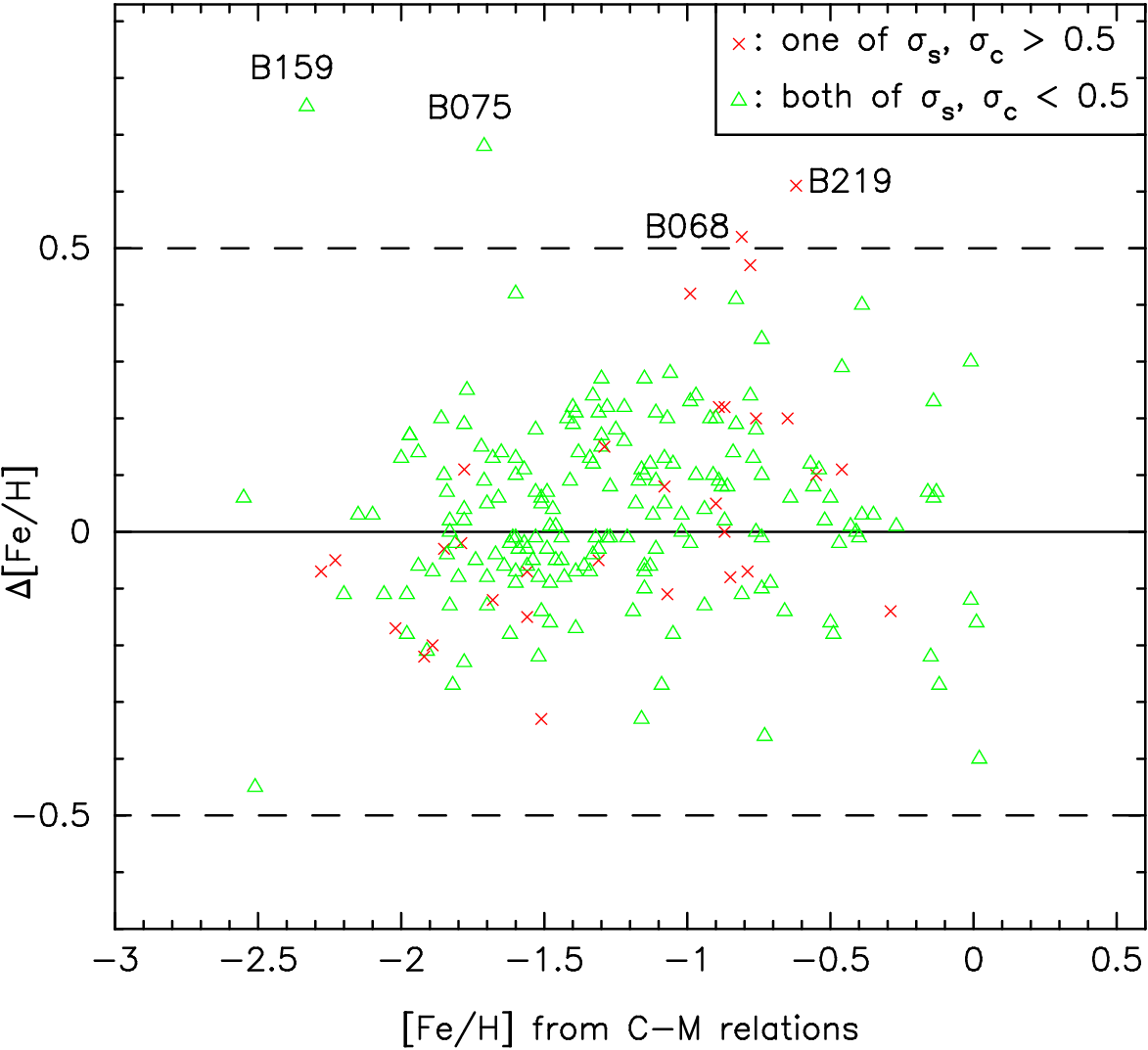}}}
\caption{Comparison of spectroscopic and colour-derived
metallicities for M31 GCs with spectroscopic data; $\sigma_{\rm s}$
indicates the uncertainty in the spectroscopic metallicity and
$\sigma_{\rm c}$ is based on the C-M relations.} \label{fig7}
\end{center}
\end{figure}

\begin{figure}
\begin{center}
\resizebox{\hsize}{!}{\rotatebox{-90}{\includegraphics{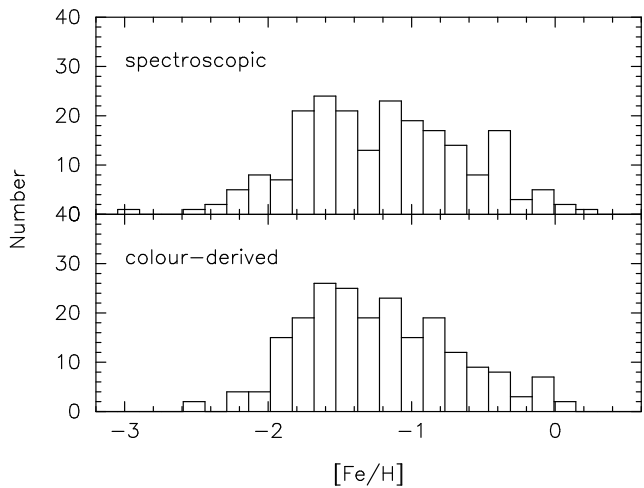}}}
\caption{Comparison of spectroscopic and colour-derived metallicity
distributions.} \label{fig8}
\end{center}
\end{figure}

In this section, we will test the process outlined in Section 4.1 by
applying it to the clusters with spectroscopic metallicities; this
includes all GCs for which the metallicities can be determined based
on the C-M relation fits. The results of our comparison are shown in
Fig. \ref{fig7}; the mean metallicity offset (spectroscopic minus
colour-derived metallicity) is $0.039\pm0.022$ dex. From Fig.
\ref{fig7} we can see that there is no evidence of a bias in the
prediction of the metallicities. The offsets for all clusters are
less than 0.7 dex, and the offsets for 4 clusters are greater than
0.5 dex. Since some of these metallicity differences are
substantial, we have carefully double checked our data and results.
In fact, the large spread in metallicity space is also evident from
fig. 10 of \citet{bh00}. However, upon close examination of the
data, we found that there are 4 objects that should not be included
in the calculation of the offset between spectroscopic and
colour-derived metallicities, i.e. objects B068, B075, B159 and
B219. For B068, we determined 14 colour-derived metallicities, and
these colour-derived metallicities have a small scatter. Their mean
value is ${\rm [Fe/H]=-0.81\pm0.03}$ dex. However, the spectral
metallicity is ${\rm [Fe/H]=-0.29\pm0.59}$ dex; we suspect that the
spectral metallicity determination may be incorrect. For B075, there
are only three colour-derived metallicities, the mean value of which
is ${\rm [Fe/H]=-1.71\pm0.08}$ dex. On the other hand, the spectral
metallicity is ${\rm [Fe/H]=-1.03\pm0.33}$ dex, so that we think
that more photometric data is needed to determine the colour-derived
metallicities more accurately. For B159, there are also only three
colour-derived metallicities. Their mean value is ${\rm
[Fe/H]=-2.33\pm0.11}$ dex, whilst the GC'sspectral metallicity is
${\rm [Fe/H]=-1.58\pm0.41}$ dex. Therefore, we also think that more
photometric data is needed to determine its colour-derived
metallicity more accurately. Finally, for B219 we determined 14
colour-derived metallicities, which exhibit a small scatter. The
mean value of these metallicities is ${\rm [Fe/H]=-0.62\pm0.05}$
dex. However, the spectral metallicity is ${\rm
[Fe/H]=-0.01\pm0.57}$ dex, while we also suspect that this spectral
metallicity may be problematic. Except for these four clusters, the
mean metallicity offset (spectroscopic minus colour-derived
metallicity) is $0.028\pm0.022$ dex, compared with $0.020\pm0.021$
found by \citet{bh00} based on a smaller GC sample. This bias in the
metallicity determination may come from large errors in either the
colour- or spectroscopically determined metallicities, or both. In
addition, the correlations between optical colours and metallicity,
which are used to determine the colour-derived metallicity for M31
clusters, are constructed based on the Galactic GCs, which may have
introduced a small (but likely insignificant) bias. In the following
analysis, we have substracted this offset from all of our
colour-derived metallicities. Fig. \ref{fig8} shows the metallicity
distributions of the spectroscopic and colour-derived samples (cf.
Fig. \ref{fig7}). Here, we now use a Kolmogorov-Smirnov (KS) test to
demonstrate whether the two distributions in Fig. \ref{fig8} are the
same. We determined a value of $D_{\rm max}=0.061$ for these two
samples (which do not include the four objects noted above); $D_{\rm
max}$ is defined as the maximum value of the absolute difference
between two cumulative distribution functions. The probability of
obtaining a value of $D_{\rm max}=0.061$ is 80.8\%. It is clear
that the KS test supports the similarity of the metallicity
distributions of the spectroscopic and colour-derived samples.

\begin{table*}
\caption{New metallicity estimates for M31 GCs and GC candidates
without spectroscopic observations.}
\label{t7.tab}
\begin{center}
\begin{tabular}{lrlrlrlrlr}
\hline
\hline \multicolumn{1}{c}{Name}& \multicolumn{1}{c}{$\rm [Fe/H]$} & \multicolumn{1}{c}{Name}& \multicolumn{1}{c}{$\rm [Fe/H]$} & \multicolumn{1}{c}{Name}& \multicolumn{1}{c}{$\rm [Fe/H]$} & \multicolumn{1}{c}{Name}& \multicolumn{1}{c}{$\rm [Fe/H]$} & \multicolumn{1}{c}{Name} & \multicolumn{1}{c}{$\rm [Fe/H]$} \\
\hline
B290    & $ -1.07\pm 0.07$ &BA21    & $ -2.51\pm 0.11$ &B412    & $ -0.80\pm 0.02$ &B413    & $ -1.57\pm 0.08$ &B134D   & $ -2.41\pm 0.23$ \\
B291    & $ -1.12\pm 0.04$ &B138D   & $ -0.36\pm 0.04$ &B140D   & $ -1.57\pm 0.12$ &B141D   & $ -1.07\pm 0.08$ &B142D   & $ -2.59\pm 0.24$ \\
B144D   & $ -1.62\pm 0.09$ &B148D   & $ -1.93\pm 0.21$ &B149D   & $ -2.21\pm 0.25$ &B150D   & $ -2.52\pm 0.07$ &B416    & $ -1.34\pm 0.07$ \\
B152D   & $ -2.55\pm 0.09$ &B418    & $ -1.19\pm 0.10$ &B154D   & $ -0.53\pm 0.83$ &B156D   & $ -2.58\pm 0.14$ &B420    & $ -0.63\pm 0.07$ \\
B157D   & $ -0.09\pm 0.08$ &B422    & $ -1.97\pm 0.18$ &B162D   & $ -2.53\pm 0.17$ &B163D   & $  0.20\pm 0.06$ &B424    & $ -2.09\pm 0.10$ \\
B165D   & $ -1.05\pm 0.13$ &B166D   & $ -1.02\pm 0.08$ &B167D   & $ -2.34\pm 0.08$ &B427    & $ -1.57\pm 0.08$ &B169D   & $ -0.10\pm 0.04$ \\
B428    & $ -2.04\pm 0.09$ &B172D   & $ -2.51\pm 0.12$ &B175D   & $  0.31\pm 0.09$ &B177D   & $ -1.32\pm 0.01$ &B433    & $ -2.45\pm 0.14$ \\
B435    & $ -1.76\pm 0.12$ &B178D   & $ -0.90\pm 0.24$ &B309    & $ -2.03\pm 0.26$ &B181D   & $ -2.21\pm 0.20$ &B438    & $ -2.42\pm 0.31$ \\
B186D   & $ -2.08\pm 0.14$ &B440    & $ -0.41\pm 0.09$ &B003    & $ -2.08\pm 0.07$ &B188D   & $ -1.84\pm 0.14$ &B325    & $ -1.77\pm 0.08$ \\
B192D   & $ -0.46\pm 0.03$ &B330    & $ -1.98\pm 0.06$ &B004D   & $ -1.17\pm 0.05$ &B194D   & $ -1.97\pm 0.12$ &B447    & $ -1.73\pm 0.12$ \\
B244    & $ -1.50\pm 0.21$ &B014    & $ -0.55\pm 0.13$ &B197D   & $ -0.69\pm 0.05$ &B450    & $ -0.89\pm 0.05$ &B022    & $ -1.64\pm 0.07$ \\
B339    & $ -0.90\pm 0.04$ &B202D   & $ -0.74\pm 0.04$ &B020D   & $ -0.76\pm 0.08$ &B032    & $ -1.09\pm 0.03$ &B456    & $ -1.71\pm 0.29$ \\
B203D   & $ -2.03\pm 0.08$ &B457    & $ -1.60\pm 0.21$ &B204D   & $ -1.18\pm 0.10$ &B025D   & $ -1.28\pm 0.10$ &B205D   & $ -2.35\pm 0.21$ \\
B207D   & $ -0.27\pm 0.08$ &B052    & $  0.12\pm 0.17$ &B060    & $ -1.87\pm 0.06$ &B062    & $ -0.47\pm 0.11$ &B067    & $ -2.55\pm 0.04$ \\
B257    & $ -2.05\pm 0.82$ &B461    & $ -2.56\pm 0.07$ &B070    & $ -1.66\pm 0.10$ &B078    & $ -0.56\pm 0.15$ &B079    & $ -0.85\pm 0.03$ \\
B345    & $ -1.40\pm 0.06$ &B462    & $ -2.28\pm 0.34$ &B084    & $ -0.76\pm 0.07$ &B346    & $ -1.70\pm 0.07$ &B347    & $ -1.71\pm 0.03$ \\
B348    & $ -1.38\pm 0.07$ &B463    & $ -1.46\pm 0.18$ &B099    & $ -1.03\pm 0.06$ &B100    & $ -2.21\pm 0.10$ &B101    & $ -1.17\pm 0.02$ \\
NB46    & $ -1.48\pm 0.03$ &NB70    & $ -2.48\pm 0.04$ &B464    & $ -0.44\pm 0.09$ &B111    & $ -1.50\pm 0.03$ &B260    & $ -0.36\pm 0.10$ \\
NB64    & $ -2.12\pm 0.46$ &B118    & $ -1.64\pm 0.10$ &B351    & $ -1.60\pm 0.05$ &NB25    & $ -0.31\pm 0.08$ &B123    & $ -1.58\pm 0.04$ \\
B128    & $ -0.92\pm 0.05$ &NB50    & $ -2.23\pm 0.19$ &B136    & $ -2.39\pm 0.08$ &B217D   & $ -2.36\pm 0.04$ &B266    & $ -2.80\pm 0.15$ \\
B220D   & $ -2.69\pm 0.13$ &B221D   & $ -1.56\pm 0.15$ &B150    & $ -0.76\pm 0.08$ &B223D   & $ -0.23\pm 0.08$ &B468    & $ -2.16\pm 0.12$ \\
B155    & $ -0.84\pm 0.03$ &B226D   & $ -2.01\pm 0.19$ &B162    & $ -0.70\pm 0.05$ &B228D   & $  0.27\pm 0.14$ &B168    & $ -0.12\pm 0.21$ \\
B169    & $ -2.56\pm 0.06$ &B172    & $ -0.87\pm 0.03$ &DAO062  & $ -2.13\pm 0.14$ &B173    & $ -1.86\pm 0.47$ &B177    & $ -0.88\pm 0.10$ \\
B181    & $ -1.10\pm 0.03$ &B231D   & $ -0.12\pm 0.07$ &B470    & $ -2.18\pm 0.23$ &B187    & $ -1.72\pm 0.04$ &B471    & $ -2.18\pm 0.09$ \\
B189    & $  0.18\pm 0.11$ &B194    & $ -1.56\pm 0.05$ &B195    & $ -1.48\pm 0.63$ &B196    & $ -1.94\pm 0.08$ &B202    & $ -1.84\pm 0.11$ \\
B361    & $ -1.61\pm 0.02$ &B237D   & $ -0.78\pm 0.26$ &G260    & $ -2.45\pm 0.06$ &B239D   & $ -1.67\pm 0.13$ &B215    & $ -1.21\pm 0.03$ \\
G268    & $ -1.36\pm 0.01$ &B243D   & $ -1.28\pm 0.07$ &B245D   & $ -2.88\pm 0.09$ &B473    & $ -2.17\pm 0.16$ &B247D   & $ -1.90\pm 0.23$ \\
B227    & $ -1.28\pm 0.08$ &B250D   & $ -0.98\pm 0.14$ &B252D   & $ -2.83\pm 0.09$ &B474    & $ -2.12\pm 0.10$ &B256D   & $ -2.37\pm 0.13$ \\
B476    & $ -0.03\pm 0.13$ &B236    & $ -1.01\pm 0.17$ &B258D   & $ -2.47\pm 0.08$ &B260D   & $ -1.46\pm 0.15$ &B478    & $ -2.69\pm 0.01$ \\
B261D   & $ -2.45\pm 0.19$ &B263D   & $ -0.85\pm 0.07$ &B286    & $ -1.67\pm 0.11$ &B479    & $ -0.36\pm 0.13$ &B266D   & $ -2.34\pm 0.10$ \\
B481    & $ -1.45\pm 0.13$ &B270D   & $ -2.28\pm 0.19$ &B273D   & $ -1.01\pm 0.03$ &B274D   & $ -0.08\pm 0.11$ &B275D   & $ -1.81\pm 0.13$ \\
B277D   & $ -0.83\pm 0.06$ &B278D   & $ -2.46\pm 0.19$ &B385    & $ -0.86\pm 0.14$ &B283D   & $ -1.55\pm 0.17$ &B288D   & $ -2.58\pm 0.23$ \\
B489    & $ -0.04\pm 0.10$ &B490    & $  0.08\pm 0.07$ &G325    & $  0.12\pm 0.12$ &B389    & $ -0.35\pm 0.08$ &B293D   & $ -2.57\pm 0.11$ \\
B295D   & $ -2.22\pm 0.13$ &B296D   & $ -0.91\pm 0.16$ &B297D   & $  0.10\pm 0.08$ &B298D   & $ -2.35\pm 0.11$ &B299D   & $ -1.91\pm 0.10$ \\
B300D   & $ -2.52\pm 0.11$ &B393    & $ -1.41\pm 0.05$ &B492    & $ -1.06\pm 0.27$ &B302D   & $ -1.30\pm 0.05$ &B304D   & $ -2.41\pm 0.12$ \\
B493    & $  1.07\pm 0.28$ &B494    & $ -1.54\pm 0.05$ &B307D   & $ -1.64\pm 0.10$ &B308D   & $ -2.43\pm 0.14$ &B495    & $ -0.35\pm 0.05$ \\
B396    & $ -1.87\pm 0.09$ &B310D   & $ -0.93\pm 0.05$ &B313D   & $ -1.00\pm 0.16$ &B314D   & $ -0.18\pm 0.06$ &B317D   & $ -2.47\pm 0.13$ \\
B319D   & $ -1.86\pm 0.17$ &B320D   & $ -2.62\pm 0.15$ &B324D   & $ -2.27\pm 0.00$ &B398    & $ -0.72\pm 0.08$ &B399    & $ -1.69\pm 0.09$ \\
B326D   & $ -0.63\pm 0.08$ &B328D   & $ -1.67\pm 0.06$ &B329D   & $ -0.24\pm 0.04$ &B330D   & $ -0.55\pm 0.06$ &B331D   & $ -1.50\pm 0.08$ \\
B332D   & $ -0.65\pm 0.09$ &B402    & $ -1.18\pm 0.06$ &B334D   & $ -0.75\pm 0.06$ &B338D   & $ -1.86\pm 0.06$ &B339D   & $ -1.51\pm 0.04$ \\
B340D   & $  0.19\pm 0.29$ &B508    & $ -2.61\pm 0.07$ &B343D   & $ -0.49\pm 0.05$ &B344D   & $ -1.40\pm 0.03$ &B345D   & $ -0.39\pm 0.12$ \\
B346D   & $  -0.37\pm0.19$ &B347D   & $  0.00\pm 0.15$ &B348D   & $ -1.16\pm 0.26$ &B349D   & $ -0.76\pm 0.26$ &        &                  \\

\hline
\end{tabular}
\end{center}
\end{table*}

\subsection{Metallicity distribution}

The metallicity distribution of the M31 clusters has been investigated
in previous studies, including \citet{hbk91}, \citet{ash93},
\citet{bh00} and \citet{per02}. With the current largest GC and GC
candidate sample at hand, we will now reanalyse the M31 GC metallicity
distribution. Including the metallicities determined based on the C-M
fits, our sample includes a total of 504 metallicities.

\begin{figure*}
\begin{center}
\resizebox{\hsize}{!}{\rotatebox{-90}{\includegraphics{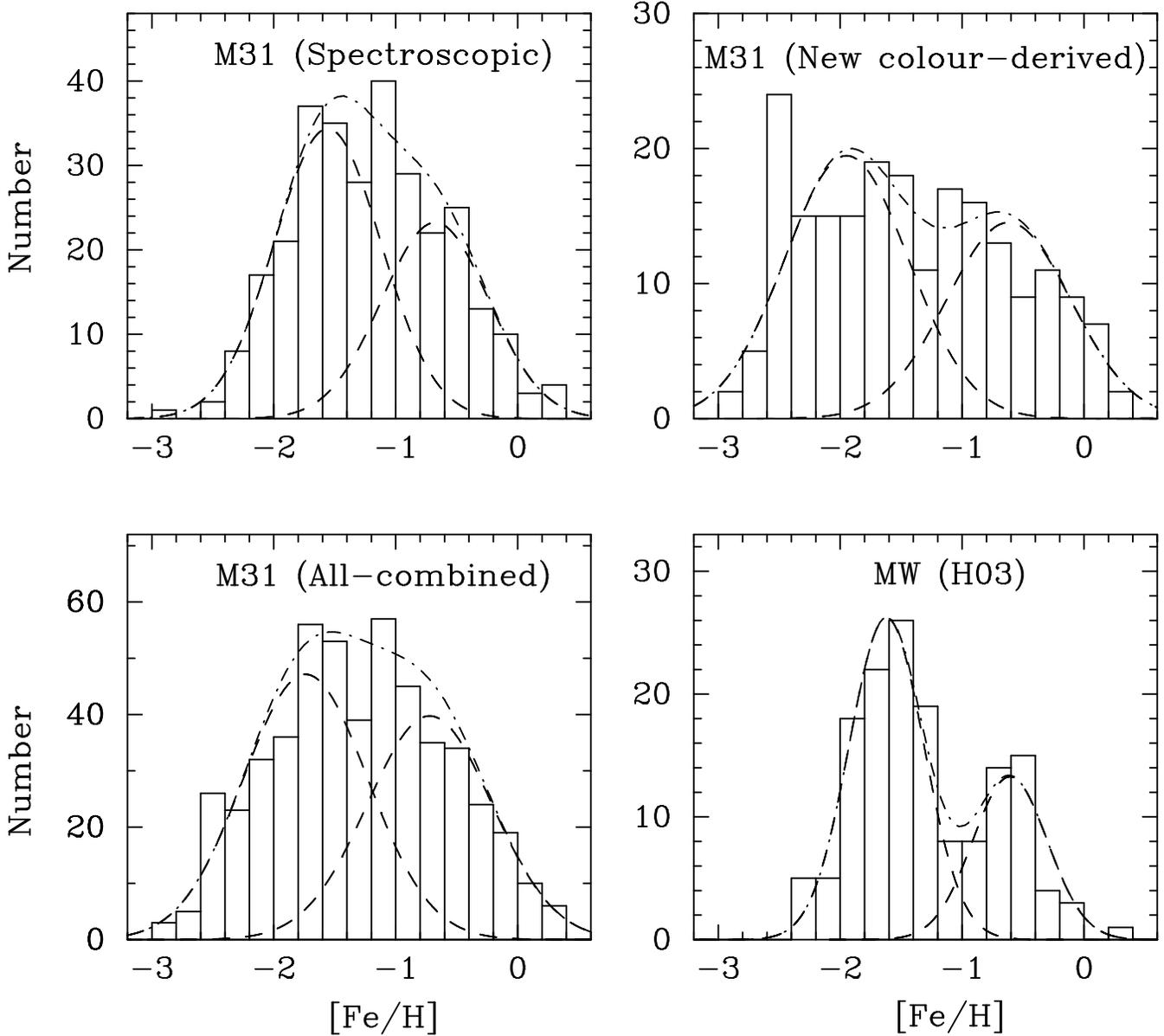}}}
\caption{Metallicity distributions and homescedastic bimodal KMM
tests of M31 GCs and GC candidates, subdivided by uncertainty, and
Galactic GCs.} \label{fig9}
\end{center}
\end{figure*}

Fig. \ref{fig9} shows the metallicity distribution of the M31 GCs
discussed in this paper, along with a similar figure for the Milky
Way GCs (from H03), for comparison. To assess whether the
combination of spectral and colour-derived metallicities has an
effect on the results, we consider three data sets in our analysis
of the metallicity distribution of the M31 objects: Set 1 contains
all objects for which metallicities have been determined from
spectroscopic observations; Set 2 contains the metallicities without
spectroscopic observations, which have been determined based on the
C-M relationships in this paper; and Set 3 contains all
metallicities from Sets 1 and 2. The mean metallicities of Sets 1,
2, and 3 are [Fe/H] = $-1.202\pm0.036$, $-1.414\pm0.057$, and
$-1.290\pm0.032$ dex, respectively, i.e. comparable to (although not
in all cases the same as) the value of $\rm [Fe/H]=-1.21\pm0.03$ dex
obtained by \citet{per02}, and also comparable to the value of $\rm
[Fe/H]=-1.298\pm0.046$ dex obtained for the Milky Way GCs (H03).
However, we point out that the mean metallicities of Sets 1 and 2
are different at the $3\sigma$ level. We investigated this result
carefully, and found that there are many more objects of which the
metallicities determined based on the C-M relationships in this
paper are lower than $\rm [Fe/H]=-2.5$, compared to the number of
GCs for which metallicities were determined from spectroscopic
observations. Generally speaking, the objects without spectroscopic
observations are fainter than those with spectroscopic observations.
Therefore, we should keep in mind that the objects for which the
metallicities have been determined based on the C-M relationships
are fainter than those with spectroscopic observations. The
photometry is more accurate for bright than for faint objects,
however. For bright clusters, i.e., those with spectroscopic
observations, the metallicities determined based on the C-M
relationships are consistent with those determined from
spectroscopic observations (see Fig. 8). In fact, \citet{bh00} did
not use the very high or low metallicity values, $\rm [Fe/H]>0.5$ or
$\rm [Fe/H]<-2.5$ dex, obtained based on the C-M relationships in
their paper when investigating the metallicity distribution. If we
exclude these very high or low metallicity values, the mean
metallicities of Sets 1 and 2 are the same. At the same time, we
emphasize that there are no reasons for not using these low or high
metallicities when investigating the metallicity distribution for
M31 GCs. The asymmetric appearance of the metallicity distribution
in Fig. \ref{fig9} suggests the possibility of bimodality. We use
the KMM algorithm \citep{McLachlan88,as94} to search for bimodality
in the metallicity distribution. The input of the KMM algorithm
includes the individual data points, the Gaussian group membership
to be fitted, and starting values for the estimated means and common
variance. In fact, for two-group homescedastic fitting, the KMM
algorithm is insensitive to these input values and converges to the
``correct'' two-group fit even when its starting points are far from
the true means and variance \citep[see details in] []{as94}. We
assumed the same variances for both groups in the metallicity
distribution; in this homoscedastic fitting the $p$-value of the
hypothesis test returned by the KMM algorithm adequately indicates
that a two-group fit is an improvement with respect to a one-group
fit. Table \ref{t8.tab} lists the parameters returned by the KMM
algorithm. It is clear that the metallicity distributions of both
data sets show strong bimodality. Thus, the KMM tests suggest that
we can consider these two distributions of metallicity bimodal at
the $\sim 100$ per cent confidence level. We also investigated the
KMM results based on three groups and on heteroscedastic two-group
fits. The results of these exercises are listed in Tables 7 and 8,
which show that three-group and heteroscedastic two-group fits are
also statistically acceptable. We point out that KMM tests assume
Gaussian distributions, which may or may not be realistic here. In
the following analysis, we investigate the metallicity distribution
for all samples in this paper, for which the homescedastic two-group
fits may be more appropriate (see Tables 6, 7 and 8). We also apply
the Dip test to confirm whether or not the metallicity distributions
of our M31 GC samples are multimodal. The Dip statistic is the
maximum difference between the empirical distribution function, and
the unimodal distribution function that minimizes the maximum
difference \citep[see details in] []{hartigan85a,hartigan85b}. The
Dip statistic measures the departure of the sample's unimodality. If
its value approaches zero, the samples are taken from a unimodal
distribution; if the Dip measure is positive, the samples come from
a multimodal distribution. The Dip values for Sets 1, 2 and 3 are
0.024, 0.025 and 0.018 with significance values of 71.2\%,
55.43\% and 66.9\%, respectively, which shows that the Dip
values support the multimodality found from the KMM tests. We note
that the Dip value for the (bimodal) metallicity distribution of the
Galactic GCs is 0.039 with significance value of 90.5\%.

\begin{table*}
\caption{Results from the KMM homescedastic bimodality tests for the metallicities of the GCs in M31 and the Milky Way.} \label{t8.tab}
\begin{center}
\begin{tabular}{ccccccccc}
\hline
\hline Data Set& $\overline{\rm [Fe/H]}$ &$\sigma_{\rm [Fe/H]} $&
$\overline{\rm [Fe/H]_{1}}$ & $\overline{\rm [Fe/H]_{2}}$ &
$\sigma_{\rm [Fe/H]} $ & $n_1$ & $n_2$ & $p$ \\
\hline
 1 &$ -1.202\pm0.036$ &0.617& $-1.558 $&$-0.676 $&0.438  &182 &113 & 0.032 \\
 2 &$ -1.414\pm0.057$ &0.828& $-1.948 $&$-0.619 $&0.508  &124 &85  & 0.000 \\
 3 &$ -1.290\pm0.032$ &0.719& $-1.740 $&$-0.722 $&0.510  &284 &220 & 0.004 \\
 MW&$ -1.298\pm0.046$ &0.564& $-1.620 $&$-0.608 $&0.306  &101 &47  & 0.000 \\
\hline
\end{tabular}
\end{center}
\end{table*}

\begin{table*}
\caption{Results from the KMM heteroscedastic bimodality tests for
the metallicities of the GCs in M31.} \label{t9.tab}
\begin{center}
\begin{tabular}{cccccccc}
\hline
\hline Data Set& $\overline{\rm [Fe/H]_{1}}$ & $\overline{\rm [Fe/H]_{2}}$ &
$\sigma_{1,\rm [Fe/H]} $ & $\sigma_{2,\rm [Fe/H]} $ &  $n_1$ & $n_2$ & $p$ \\
\hline
 1 & $-1.487 $&$-0.574 $&0.466 &0.401   &207 &88 & 0.145 \\
 2 & $-2.424 $&$-1.192 $&0.180 &0.743   &44  &165  & 0.000 \\
 3 & $-1.794 $&$-0.786 $&0.488 &0.534   &254 & 250 & 0.022 \\
\hline
\end{tabular}
\end{center}
\end{table*}

\begin{table*}
\caption{Results from the KMM homescedastic trimodality tests for the metallicities of the GCs in M31.} \label{t10.tab}
\begin{center}
\begin{tabular}{ccccccccc}
\hline
\hline Data Set& $\overline{\rm [Fe/H]_{1}}$ & $\overline{\rm [Fe/H]_{2}}$ &
$\overline{\rm [Fe/H]_{3}}$ &$\sigma_{\rm [Fe/H]} $ &  $n_1$ & $n_2$ &$n_3$
& $p$ \\
\hline
 1 & $-1.733 $&$-1.100 $& $-0.462 $& 0.374 & 120 & 114 &  61 & 0.128 \\
 2 & $-2.141 $&$-1.159 $& $-0.266 $& 0.399 &  92 &  78 &  39 & 0.000 \\
 3 & $-1.985 $&$-1.206 $& $-0.434 $& 0.419 & 162 & 237 & 105 & 0.012 \\
\hline
\end{tabular}
\end{center}
\end{table*}

\subsection{Spatial distribution}

In the previous section, we investigated the metallicity
distribution of the M31 GCs based on the KMM test. The KMM test
allows us to distinguish between the metal-poor and metal-rich
subsamples, i.e., the KMM test gives {\rm [Fe/H]=-1.18} dex as the
dividing line between the metal-poor and metal-rich GCs. However,
there are about 54 objects that exhibit intermediate probabilities
of membership in both groups ($0.5<prob.<0.6$). Since it is
difficult to decide unequivocally which group these objects belong
to, we simply adopted the dividing line between metal-poor and
metal-rich GCs from the KMM test.

\begin{figure*}
\begin{center}
\includegraphics[angle=-90,width=120mm]{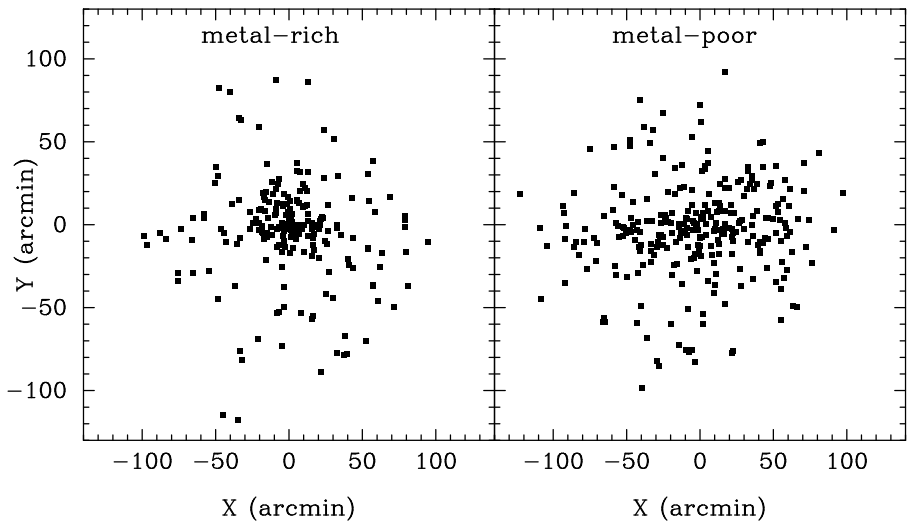}
\caption{Spatial distribution of the metal-rich and metal-poor
GCs in M31.} \label{fig10}
\end{center}
\end{figure*}

\begin{figure}
\begin{center}
\resizebox{\hsize}{!}{\rotatebox{-90}{\includegraphics{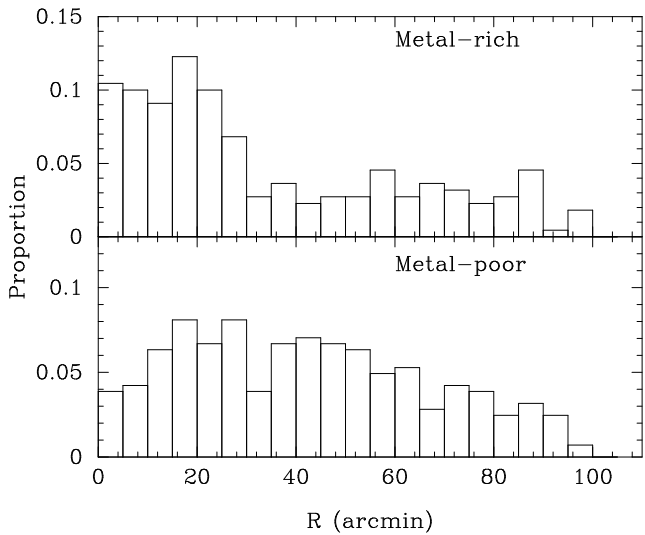}}}
\caption{Radial distribution of the metal-rich and metal-poor GCs and
GC candidates in M31.} \label{fig11}
\end{center}
\end{figure}

Figure \ref{fig10} shows the projected spatial distributions of the
metal-poor and metal-rich GCs in M31. Using Eqs. (7) and (8), we
obtained the distances to our sample clusters from the centre of
M31. From Fig. \ref{fig10}, it is clear that the metal-rich GCs in
M31 are more centrally concentrated, consistent with what was found
by \citet{hbk91} and \citet{per02}. The metal-poor GCs appear to
occupy a more extended halo, although also with a general
concentration following the outline of the M31 disc. The latter may
have been caused by selection biases, i.e., from Fig. 4 we can
see that there are more clusters observed along the major axis than
the minor axis. Fig. \ref{fig11} shows the histograms of the
metal-poor and metal-rich populations. A notable shortage of
metal-poor clusters in the innermost radial bins can clearly be
seen, which is also consistent with the results of \citet{per02}. In
the Milky Way, the metal-rich GCs reveal significant rotation and
have historically been associated with the thick-disc system
\citep{zinn85,taft89}; however, other studies
\citep{fw82,minniti95,patr99,forbes01} have suggested that
metal-rich GCs within $\sim 5$ kpc of the Galactic Centre are more
appropriately associated with the Milky Way's bulge and/or bar. In
M31, \citet{ew88} showed that the metal-rich clusters constitute a
more highly flattened system than the metal-poor ones, and appear to
have disc-like kinematics; \citet{hbk91} showed that the metal-rich
GCs are preferentially located close to the galactic centre.
\citet{hbk91} also showed that the distinction between the rotation
of the metal-rich and metal-poor clusters is most apparent in the
inner 2 kpc. Therefore, they concluded that the metal-rich clusters
in M31 appear to form a central rotating disc system. \citet{per02}
performed a more comprehensive investigation into the kinematics of
the M31 cluster system. They showed that the metal-rich M31 GCs
appear to constitute a distinct kinematic subsystem that
demonstrates a centrally concentrated spatial distribution with a
high rotation amplitude, but that does not appear significantly
flattened, consistent with a bulge population. \citet{sbkhp02}
performed a maximum-likelihood kinematic analysis of 166 M31
clusters taken from \citet{bh00} and found that the most significant
difference between the rotation of the metal-rich and metal-poor
clusters occurs at intermediate projected galactocentric radii.
Particularly, \citet{sbkhp02} presented a potential thick-disc
population among M31's metal-rich GCs.

\subsection{Metallicity gradient}

The presence or absence of a radial trend in the metallicity of a GC
sample is an important test of galaxy formation theories
\citep{bh00}. In the \citet{eggen62} galaxy formation scenario, the
halo stars and GCs should show large-scale metallicity gradients
\citep{eggen62,bh00}; however, in the \citet{sz78} scenario the
expected metallicity distribution is more homogeneous. For the Milky
Way, \citet{taft89} provided some evidence that metallicity gradients
with both distance from the Galactic plane and distance from the
Galactic Centre are present in the disc cluster system. For M31, there
are some inconsistent conclusions, e.g., \citet{van69} showed that
there is little or no evidence for a correlation between metallicity
and projected radius, but most of his clusters were located inside a
radius of 50 arcmin; however, some authors \citep[see,
e.g.,][]{hsv82,sha88,hbk91,per02} showed that there is evidence for a
weak but measurable metallicity gradient as a function of projected
radius. \citet{bh00} confirmed the latter result based on their large
sample of spectral and colour-derived metallicities.

\begin{figure}
\begin{center}
\resizebox{\hsize}{!}{\rotatebox{-90}{\includegraphics{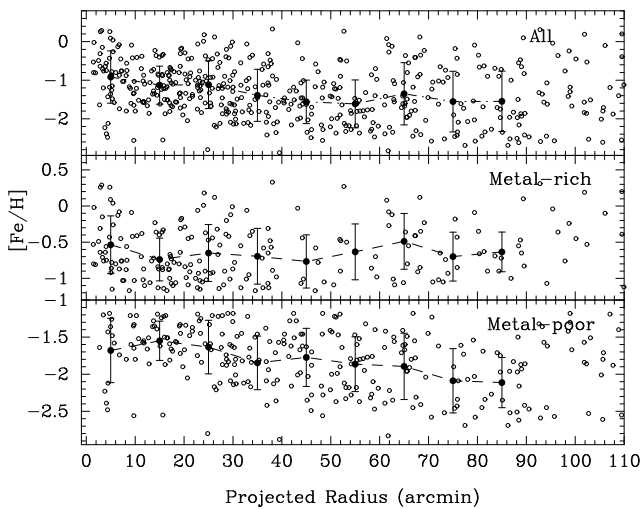}}}
\caption{Metallicity distribution versus projected radius for the
entire and the metal-poor populations of the M31 GCs and GC
candidates.}
\label{fig12}
\end{center}
\end{figure}

In Fig. \ref{fig12}, we show the metallicity of the M31 GCs as a
function of galactocentric radius based on our large cluster sample.
It is clear that the dominant feature of this diagram is the scatter
in metallicity at any radius. However, it is also true that a trend
of decreasing metallicity with increasing galactocentric distance
exists, for both the metal-poor and the entire population. The
slopes of the metal-poor subsample and for the entire sample are
$-0.006\pm 0.001$ and $ -0.007\pm0.002$ dex arcmin$^{-1}$,
respectively, while for metal-rich sample, it is $0.000\pm 0.001$
dex arcmin$^{-1}$. The latter can certainly be considered as no
metallicity gradient. In order to show this, we display the mean
metallicity binned in 10 arcmin intervals in galactocentric radius.
We can see that, within $\sim 90$ arcmin, the mean metallicity
decreases with galactocentric radius for both the metal-poor and for
the entire population. The error bars represent the standard
deviations of the means. These results are in good agreement with
\citet{per02} and \citet{hbk91}. Therefore, we can conclude that
simple smooth, pressure-supported collapse models of galaxies by
themselves are unlikely to fit M31.

\subsection{Metallicity versus intrinsic magnitude}

The (possible) correlation between cluster mass (or luminosity) and
metallicity is important in GC formation theory. It is generally
believed that if self-enrichment is important in GCs, the most massive
clusters could retain their metal-enriched supernova ejecta, so that
the metal abundance should increase with cluster mass; the opposite is
true if cooling from metals determines the temperature in the
cluster-forming clouds \citep{bh00}. The possible self-enrichment of
GCs has been studied in detail in some aspects \citep[see details
in][]{sbsb06}. However, the model of GC self-enrichment developed by
\citet{pj99} is particularly interesting in this context. In this
model, cold and dense clouds embedded in the hot proto-galactic medium
are assumed to be the progenitors of galactic halo GCs.  Based on this
model, \citet{pg01} suggested that the most metal-rich proto-GCs are
the least massive ones.

The {\sl HST} provides a unique tool to study GCs in external
galaxies. Recently, using the ACS onboard the {\sl HST},
\citet{harris06}, \citet{mieske06} and \citet{sbsb06} found that in
giant ellipticals -- such as M87, NGC 4649 and NGC 7094 (although not
in NGC 4472) -- luminous blue GCs reveal a trend of having redder
colours, such that more massive GCs are redder (more metal-rich). This
trend is referred to as the ``blue tilt'' \citep[see
also][]{strader06}.  This blue tilt has been interpreted as a result
of self-enrichment \citep{sbsb06}. \citet{sbsb06} speculatively
suggested that these GCs once possessed dark matter haloes.
\citet{spitler06} subsequently found that this blue tilt is also
present in the Sombrero spiral galaxy (NGC 4594) and may extend to
less luminous GCs with a somewhat shallower slope than derived by
\citet{harris06} and \citet{sbsb06}. As \citet{spitler06} pointed out,
the Sombrero galaxy provides the first example of this trend in a
spiral galaxy, and in a galaxy found in a low-density galaxy
environment. However, in these ACS studies, the metal-rich (redder)
GCs did not show a corresponding trend \citep[see
also][]{bekki07}. Based on high-resolution cosmological simulations
including GCs, \citet{bekki07} investigated the formation processes
and physical properties of GC systems in galaxies, and found that
luminous metal-poor clusters would develop a correlation between
luminosity and metallicity if they originated from the nuclei of
low-mass galaxies at high redshift. In fact, in the simulations of
\citet{bekki07}, the ``simulated blue tilts'' come from the assumption
that luminous metal-poor clusters originate from the stellar galactic
nuclei of the more massive nucleated galaxies exhibiting a
luminosity-metallicity relation. It is therefore evident that, in
\citet{bekki07}, galaxies which experienced more accretion/merging
events of nucleated low-mass galaxies are more likely to show a blue
tilt.

\begin{figure}
\begin{center}
\resizebox{\hsize}{!}{\rotatebox{0}{\includegraphics{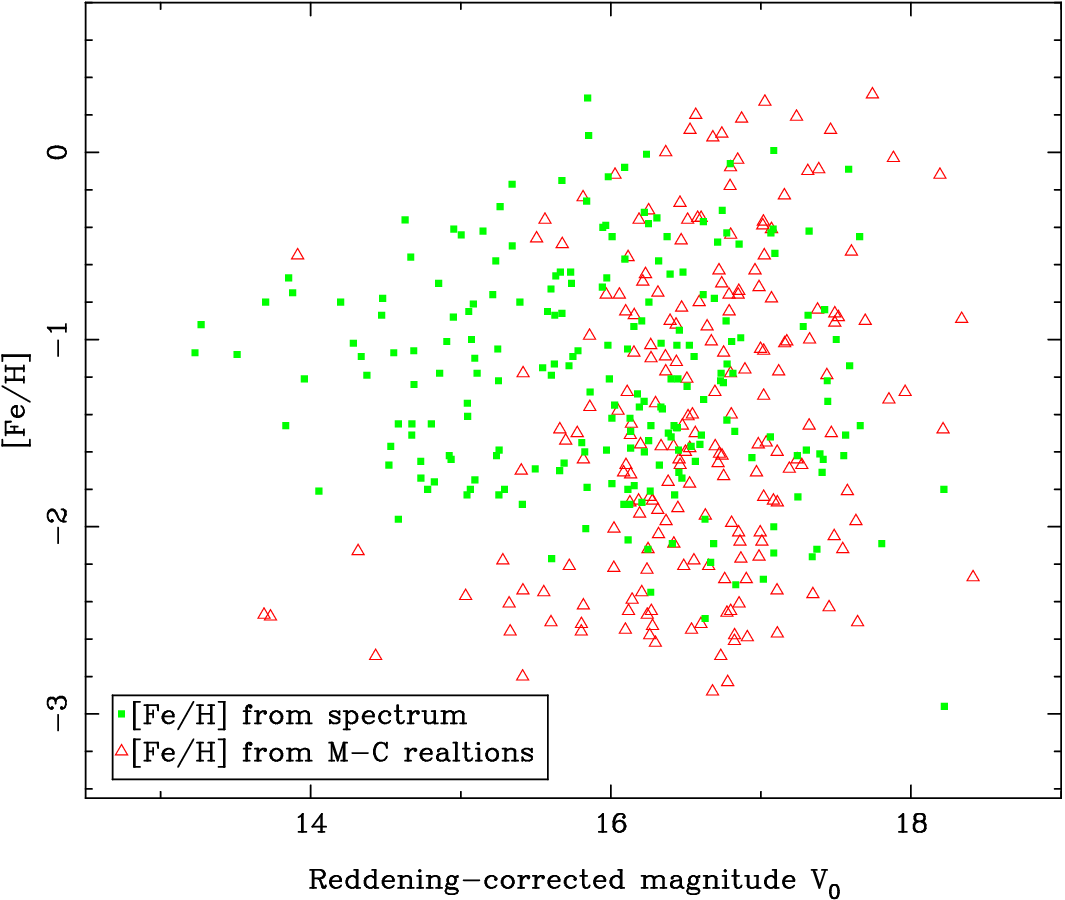}}}
\caption{Metallicity as a function of the reddening-corrected
magnitude $V_0$ for our sample GCs.} \label{fig13}
\end{center}
\end{figure}

Fig. \ref{fig13} shows the diagram of GC metallicity versus dereddened
apparent magnitude. It is clear that there is no obvious trend of
metallicity with luminosity similar to that in \citet{hbk91} and
\citet{bh00}. Least-squares fits show no evidence for a relationship
between luminosity and metallicity in our sample clusters.

\section{Discussion and Conclusions}
\label{Conclu.sec}

In this paper, we have (re-)determined the reddening values for 443
clusters and cluster candidates in M31, as well as metallicities for
209 sample objects without spectroscopic observations. We have
followed the methods described by \citet{bh00}, who found that the M31
and Galactic extinction laws are the same within the observational
errors, and that the M31 and Galactic GC C-M relations are also
consistent with each other. The sample of spectroscopic and
photometric data used in this paper is the newest and largest to
date. The spectroscopic data were obtained from the most recent
references currently available and the photometric data are from the
most comprehensive catalogue of M31 clusters available at present,
which includes 337 confirmed GCs and 688 GC candidates. Using the
metallicities of the largest sample of clusters and cluster candidates
at hand, we studied the properties of the M31 clusters. Our main
conclusions are summarised below:

\begin{enumerate}
\item The reddening distribution shows that slightly more than 50 per
cent of the GCs suffer from a reddening of less than $E(B-V)=0.2$ mag,
and the mean value is $E(B-V) = 0.28_{-0.14}^{+0.23}$ mag. The spatial
distribution of $E(B-V)$ indicates that the reddening on the
northwestern side of the M31 disc is greater than that on the
southeastern side, which is consistent with the conclusion that the
northwestern side in nearer to us.

\item The metallicity distribution of the M31 GCs is bimodal with
peaks at $\rm {[Fe/H]}\approx -1.7$ and $-0.7$ dex.

\item The diagram of metallicities as a function of radius from the
M31 centre shows a metallicity gradient for the metal-poor GCs, but no
such gradient for the metal-rich GCs.

\item The metal-rich clusters appear to constitute a centrally
concentrated spatial distribution; however, the metal-poor clusters
tend to be less spatially concentrated.

\item There is no correlation between luminosity and metallicity among
our M31 sample clusters, which indicates that self-enrichment is
indeed unimportant for cluster formation in M31.
\end{enumerate}

We reiterate that in using the method of \citet{bh00}, there are two
major unavoidable assumptions (acknowledged by these authors),
i.e. that in the Milky Way and in M31 both the extinction law and the
intrinsic colours of the GCs are the same. The latter assumption seems
reasonable, since there is no evidence that GCs in different galaxies
have different intrinsic colours. Regarding the former assumption,
there is inconsistent evidence as to whether or not this is a valid
assumption. For example, \citet{wk88} found that the extinction law in
M31 is very similar to that in the Milky Way, by analysing the two
major dust lanes on the near side of M31; however, several studies
have suggested that the reddening in M31 appears to be peculiar: with
$E(U-B)/E(B-V)=1.01\pm0.11$ \citep{ir85} and $E(U-B)/E(B-V)\sim 0.5$
\citep{Massey95}, compared to 0.72 for the same ratio in the Milky
Way. Based on a large sample of GCs with optical and near-infrared
photometric data, \citet{bh00} demonstrated that the $U$- and $K$-band
extinction curve of M31 is consistent with that of the Milky Way, with
total-to-selective extinction coefficient $R_V=3.1$. In fact, the
former assumption is plausible because in the M31 disc the composition
and size distribution of the large normal grains which dominate the
dust mass may be similar to those in the Milky Way \citep[see for
details,][]{xh96}.

As an example, we will discuss in some detail the reddening value of
the M31 GC B037 (a.k.a. 037-B327), which is known to be an extremely
red object. There are a few references that discuss this GC, including
\citet{bk02}, \citet{Ma06}, \citet{ma06} and \citet{cohen06}.
\citet{km60} first noticed an extremely red colour in photographic
($P$) and visual ($V$) bands for B037, and determined its absorption
to be $A_V=3.90$ mag. Based on the photometric data for M31 star
clusters in $U, B$, and $V$ of \citet{ve62}, \citet{ve62b} studied the
reddening values for these objects and found that B037 was the most
highly reddened in his sample, with $E(B-V)=1.28$ mag ($A_V=4.10$
mag). \citet{Crampton85} calibrated $(B-V)_{\rm 0}$ as a function of
spectroscopic slope parameter $S$ of the continuum between $\sim 4000$
and 5000 \AA, and then determined the intrinsic colours for about 40
GCs and GC candidates, including B037. \citet{Crampton85} presented a
reddening value for B037 of $E(B-V)=1.48$ mag. Armed with a large
database of multicolour photometry, \citet{bh00} determined the
reddening value for each individual M31 GC, including B037, using the
correlations between optical and infrared colours and metallicity
based on various ``reddening-free'' parameters, and derived
$E(B-V)=1.38\pm0.02$ mag for B037. Using spectroscopic metallicities
to predict the intrinsic colours, \citet{bk02} rederived the reddening
value for this GC, $E(B-V)=1.30\pm0.04$ mag. Recently, \citet{Ma06}
determined the reddening and age of the B037 by comparing multicolour
photometry with theoretical stellar population synthesis models. The
reddening towards B037 determined by \citet{Ma06} is
$E(B-V)=1.360\pm0.013$ mag. The reddening value for B037 determined in
this paper is $E(B-V)=1.21\pm0.03$ mag. It is clear that the
consistent reddening values for B037 from different references confirm
that this cluster suffers from very large extinction. In fact,
\citet{ma06} showed the dust lane across the face of the cluster using
an {\sl HST}/ACS image, which may partially account for its very large
reddening value \citep[see also][]{cohen06}.

\section*{Acknowledgments}
We would like to thank the referee, Terry Bridges, for providing
rapid and thoughtful report that helped improve the original
manuscript greatly. This work has been supported by the Chinese
National Natural Science Foundation Nos 10473012, 10573020,
10633020, 10673012 and 10603006; and by National Basic Research
Program of China (973 Program) No. 2007CB815403. RdG acknowledges
partial financial support from the Royal Society in the form of a
UK-China International Joint Project.

\end{document}